\documentclass[reprint,amsmath,amssymb,aps,prd,nofootinbib,floatfix]{revtex4-1}

\usepackage{hyperref}
\usepackage{amssymb,amsmath,amsfonts}
\usepackage{mathrsfs}
\usepackage{dsfont} 
\usepackage{graphicx,bm}
\usepackage{multirow}
\usepackage{color}
\usepackage{mathtools}
\graphicspath{{./figures/}}
\usepackage{orcidlink}  

\usepackage{dcolumn}% Align table columns on decimal point
\usepackage{array}

\def\u{\text{\scalebox{0.8}{$\uparrow$}}}
\def\d{\text{\scalebox{0.8}{$\downarrow$}}}
\newcommand{\Msun}{\mathrm{M}_\odot}

\allowdisplaybreaks

\begin{document}

\title{Observability of dynamical tides in merging eccentric neutron star binaries}

\author{J{\'a}nos Tak{\'a}tsy\,\orcidlink{0000-0002-2657-5094}}
\email{takatsy.janos@wigner.hun-ren.hu}
\affiliation{Niels Bohr International Academy, The Niels Bohr Institute, Blegdamsvej 17, 2100 Copenhagen, Denmark}
\affiliation{Institute for Particle and Nuclear Physics, HUN-REN Wigner Research Centre for Physics, Konkoly–Thege Miklós út 29-33, 1121 Budapest, Hungary}
\author{Bence Kocsis\,\orcidlink{0000-0002-4865-7517}}
\affiliation{
Rudolf Peierls Centre for Theoretical Physics, University of Oxford, Clarendon Laboratory, Parks Road, Oxford, OX1 3PU, UK
}
\affiliation{
St Hugh’s College, University of Oxford, St Margaret’s Rd, Oxford, OX2 6LE, UK
}
\author{Péter Kovács\,\orcidlink{0000-0003-3735-7620}}
\affiliation{Institute for Particle and Nuclear Physics, HUN-REN Wigner Research Centre for Physics, Konkoly–Thege Miklós út 29-33, 1121 Budapest, Hungary}

\begin{abstract}
While dynamical tides only become relevant during the last couple of orbits for circular inspirals, orbital eccentricity can increase their impact during earlier phases of the inspiral by exciting tidal oscillations at each close encounter. We investigate the effect of dynamical tides on the orbital evolution of eccentric neutron star binaries using post-Newtonian numerical simulations and construct an analytic stochastic model that reproduces the numerical results. Our study reveals a strong dependence of dynamical tides on the pericenter distance, with the fractional energy transferred to dynamical tides over that dissipated in gravitational waves (GWs) exceeding $\sim1\%$ at separations $r_\mathrm{p}\lesssim50$~km for large eccentricities. We demonstrate that the effect of dynamical tides on orbital evolution can manifest as a phase shift in the GW signal. We show that the signal-to-noise ratio of the GW phase shift can reach the detectability threshold of 8 with a single aLIGO detector at design densitivity for eccentric neutron star binaries at a distance of $40$~Mpc. This requires a pericenter distance of $r_\mathrm{p0}\lesssim68$~km ($r_\mathrm{p0}\lesssim76$~km) at binary formation with eccentricity close to 1 for a reasonable tidal deformability and f-mode frequency of 500 and $1.73$~kHz (700 and $1.61$~kHz), respectively. The observation of the phase shift will enable measuring the f-mode frequency of neutron stars independently from their tidal deformability, providing significant insights into neutron star seismology and the properties of the equation of state. We also explore the potential of distinguishing between equal-radius and twin-star binaries, which could provide an opportunity to reveal strong first-order phase transitions in the nuclear equation of state.
\end{abstract}

\maketitle

\section{Introduction}
\label{sec:intro}

Since the first direct observation of a binary black hole (BH) inspiral with gravitational waves (GWs) in September 2015 \cite{LIGOScientific:2016aoc}, the LIGO--Virgo--Kagra (LVK) Collaboration \cite{PhysRevD.93.112004,PhysRevLett.123.231108,KAGRA:2020tym} announced the detection of almost 100 GW signals produced by inspiraling compact object binaries \cite{KAGRA:2021vkt}, and the ongoing observing run already produced more than 117 additional significant candidate GW events\footnote{see \url{https://gracedb.ligo.org/superevents/public/O4/} for the list of candidates}. These events are dominated by binary BH mergers, however, some of them are attributed to mergers of binary neutron star (NS) and NS--BH systems. Particularly interesting was the GW event of a binary NS merger, detected in August 2017, GW170817 \cite{LIGOScientific:2017vwq}. Together with the associated gamma-ray burst and kilonova signals they were the first example of multi-messenger astronomy with GWs \cite{LIGOScientific:2017ync,LIGOScientific:2017zic}. This multi-messenger observation provided a rich laboratory for standard siren measurement of the cosmic expansion history \cite{LIGOScientific:2017adf}, the investigation of the equation of state (EoS) of strongly interacting matter through tidal deformability measurement \cite{LIGOScientific:2018hze}, as well as the inference on the origin of r-process elements \cite{Kasen:2017sxr}.

In particular, the measurement of the tidal deformability provided strong constraints on the nuclear EoS, since it excluded the stiffest equations of state by putting an upper bound on the radius of NSs with a mass of $1.4~\Msun$ \cite{Annala:2017llu,Most:2018hfd,De:2018uhw,Tews:2018iwm}. NSs, in general, are vital sources of information about the EoS of strongly interacting matter at low temperature and intermediate densities. Under such conditions, none of our microscopic models are predictive. At low densities, below nuclear saturation, the EoS is well established. Here, two- and three-body nuclear interactions are determined by experimental data mostly based on nucleon-nucleon scattering and the properties of light nuclei (e.g. \cite{Akmal1998,Wiringa1988}). Hadron resonance gas models and Chiral Effective Field Theory provide different but robust ways to determine the low-density EoS and to estimate uncertainties \cite{Huovinen2009,Bazavov2012,Lynn2015,Tews2018}. According to state-of-the-art calculations, the uncertainties of the nuclear EoS above $\sim1.1\,n_0$ become increasingly significant, with $n_0$ being the baryon number density at nuclear saturation \cite{Tews2012,Tews2019}. On the opposite side of the phase diagram, at very high densities, due to the asymptotic freedom, one can resort to perturbative QCD methods \cite{Kurkela2009,Mogliacci2013}. This method gives reliable results at $\mu_B\gtrsim2.5$~GeV, or equivalently $n_B\gtrsim40\,n_0$. In the intermediate density region the EoS is highly uncertain. While effective models based on the approximate global chiral symmetry of QCD can be used to estimate the EoS and the position of phase transitions, these transitions can virtually happen at any density in this uncertain region. Therefore, observations of NSs and the measurement of their properties can be important for constraining the EoS in this region. (e.g. Refs.~\cite{Annala:2017llu,Dietrich:2020efo,Huth:2021bsp,Marczenko:2022jhl,Takatsy:2023xzf}).

Since tidal deformations are sensitive to the mass, radius and internal structure of NSs, measuring the tidal deformability can be used to constrain the EoS. During a binary NS's inspiral the signature of tidal deformation in the GW signal mainly appears through a change in the phase of the waveform \cite{Flanagan:2007ix}. The dominant effect of the binary evolution is caused by adiabatic tides, where the distorted NS remains in equilibrium according to the tidal force exerted by its companion. These adiabatic tides can be characterized by a single constant for each multipole moment, the tidal deformability \cite{Love1909,Damour:2009vw,Flanagan:2007ix}. Adiabatic tides provide a good approximation in the regime where the eigenfrequencies of the tidal modes are much higher than the frequency of the acting force. However, finite frequency effects, also known as dynamical tides, can become relevant when approaching the merger of the binary system \cite{Hinderer:2016eia,Steinhoff:2016rfi,Schmidt:2019wrl,Andersson:2019dwg}. The effects of dynamical tides caused by the most dominant, fundamental-mode (f-mode) tides have been extensively studied for circular binaries using Newtonian gravity \cite{Shibata1994,Lai:1993di,Kokkotas:1995xe}, and within a time-domain effective-one-body approach \cite{Steinhoff:2016rfi,Schmidt:2019wrl}. GW models that describe the waveform in the frequency domain also provide a more efficient description from a practical point of view (see e.g. Refs.~\cite{Ajith:2007kx,Khan:2015jqa,Field:2013cfa,Bohe:2016gbl}).

While circular compact binary inspirals are expected to be more common, dynamical tides can be even more important in highly eccentric NS binary systems due to the broad spectrum of the acting tidal forces. 
During close encounters of the NSs, these tidal forces can excite the lowest-frequency f-mode oscillations of the NSs, which in turn influences the orbital evolution. Due to the broad spectrum of excitations, unlike in circular binaries, dynamical tides in eccentric binaries might become relevant well before the merger of the system, providing an opportunity to be observed \cite{Yang:2018bzx,Yang:2019kmf,Vick:2019cun,Gamba:2022mgx,Chaurasia:2018zhg}.

While field binaries can be born with high eccentricities, GW radiation efficiently damps eccentricities of binaries inspiraling from an initially large periapsis distance \cite{Peters:1964zz}, leaving nearly circular binaries by the time they reach the sensitive band of current GW detectors. However, measurable eccentricities can be produced for compact object binaries in dense stellar clusters during dynamical interactions, such as single-single gravitational-wave capture \cite{OLeary:2008myb,Kocsis:2011dr,Gondan:2017wzd,Takatsy:2018euo,Rasskazov:2019gjw,Gondan:2020svr,Zevin:2021rtf}, binary--single \cite{Samsing:2013kua,Samsing:2017xmd} or binary--binary interactions \cite{Zevin:2018kzq,Arca-Sedda+2021}, the secular evolution of hierarchical triple systems \cite{Wen:2002km,Antonini:2012ad,Antognini:2013lpa,Silsbee:2016djf,Hoang:2017fvh,Petrovich_Antonini2017,Rodriguez:2018jqu,Fragione:2019hqt,Fragione+2019}, or interactions in gaseous disks \cite{Tagawa:2020jnc,Samsing:2020tda,Fabj_Samsing2024,Li+2023,Rowan+2023,Whitehead+2024}. While theoretical predictions to form eccentric compact binaries are quite robust, their detection and eccentricity measurement are difficult, since template banks for matched-filtering GW searches typically assume quasi-circular binaries (e.g. Ref.~\cite{Messick2017,Davies:2020tsx,Aubin:2020goo}). Searches for burst-like GW signals might be effective in detecting highly eccentric sources \cite{Tiwari:2015gal,Ramos-Buades:2020eju,Cornish:2020dwh,Dalya:2020gra}, although their sensitivity is more difficult to quantify \cite{Klimenko:2015ypf}. On the other hand, recent studies have shown that eccentricities could be accurately measured using matched filtering templates if $e_\mathrm{10Hz}\gtrsim0.1$ \cite{Gondan:2017hbp,Gondan_Kocsis2019}. While the effects of finite eccentricity and spin precession on the GW signal are difficult to distinguish \cite{Romero-Shaw:2022fbf}, recent studies even suggest that some of the binary BH inspirals observed by the LVK detectors show significant support for non-negligible eccentricities \cite{Gamba:2021gap,Gayathri+2022,Romero-Shaw:2022fbf,Romero-Shaw:2022xko,Gupte:2024jfe}.\footnote{Note that Ref.~\cite{Iglesias:2022xfc} found in their analysis neglecting a possible spin-orbit misalignment that the incorporation of eccentricity was not preferred compared to the quasi-circular case, however eccentricty may be preferred for at least one observed system with spin-orbit misalignment}. Once observed, eccentricities can be used as efficient discriminators between isolated and dynamical binary formation channels (e.g. Ref.~\cite{Zevin:2021rtf}). While the merger rate of eccentric binary systems is highly uncertain, recent studies suggest that it may be significant \cite{Hoang:2017fvh,Rodriguez:2017pec,Samsing:2018ykz}, with both numerical and semi-analytical models indicating that they may account for up to $\sim10\%$ of the total binary BH merger rate in globular clusters \cite{Samsing:2017xmd,Samsing:2017oij,Rodriguez:2017pec,Zevin:2018kzq,Antonini:2019ulv,Kremer+2020,Zevin:2021rtf}.

Several previous studies also investigated various aspects of adiabatic and dynamical tides in binary systems with non-negligible eccentricities and the corresponding effects on GWs. These include analytic expressions for the energy and angular momentum transfer, as well as numerical assessments of the effect of dynamical tides on the orbital evolution \cite{Parisi:2017kgx,Yang:2018bzx,Yang:2019kmf,Vick:2019cun,Wang:2020iqj,Bernaldez:2023xoh}. Resonant excitations of tidal modes in circular and eccentric binary white dwarfs have also been the subject of multiple studies \cite{Rathore:2004gs,Fuller:2011,Burkart:2014}. Refs.~\cite{Yang:2018bzx,Yang:2019kmf} have previously derived the energy transfer between stellar oscillations and the orbital motion, which we rederive using a more convenient parameterization of the NS tides. Those papers have also calculated the GW phase shift caused by dynamical tides in the small eccentricity limit, finding that it is of the order of $\mathcal{O}(0.5)$ radians. Vick \& Lai \cite{Vick:2019cun} approached this problem from a different perspective, solving the equations of orbital motion numerically for high-eccentricity orbits. Their results suggest these phase shifts can become an order of magnitude larger for these eccentricities.

In this paper we investigate the effect of dynamical tides on the evolution of eccentric NS binary systems in more detail using realistic NS parameters and focus on the detectability of the induced GW phase shift. We run post-Newtonian numerical simulations and construct an analytic stochastic model to interpret the results for arbitrary eccentricities. In particular, our analytical model is based on the Lagrangian of Ref.~\cite{Flanagan:2007ix}, which is used to calculate the transmitted energy to tidal oscillation modes and the corresponding orbital phase shift generated during the dynamical interactions of binary NSs during each close approach. While resonant excitations of dynamical tides in NS binaries have been studied before, to our knowledge this is the first study to investigate the stochastic evolution of dynamical tides from orbit-to-orbit and the corresponding phase shift. We confirm that the analytical results and numerical calculations are in good agreement. Then, we also investigate if the effect of dynamical tides could be observed with current GW detectors by estimating the signal-to-noise ratio of the perturbation to the GW. While calculations suggest that quasi-universal relations hold between the tidal deformability and the f-mode frequency of NSs \cite{Chan:2014kua,Godzieba:2020bbz,Pradhan:2022vdf,Pradhan:2023zmg}, by measuring the effect of dynamical tides we should be able to measure these two quantities independently. Therefore, we also assess the effect of deviating from these relations. Another useful aspect of independent measurements on tidal deformabilities and f-mode frequencies is the possibility of observing inspirals of the so-called twin star binaries, i.e. NS-NS binaries with the same mass but different stellar radii. While adiabatic tides are known to be only sensitive to the mass-weighted tidal deformability of the binary in the circular case \cite{Flanagan:2007ix} with finite eccentricities only negligibly modifying these tides \cite{Bernaldez:2023xoh}, we show that dynamical tides could be used to break the degeneracy and observationally differentiate between the equal-radius and the twin star scenarios.

This paper is organised as follows. In Sec.~\ref{sec:model}, we introduce our model for eccentric binary evolution with tidal interactions, and summarize our assumptions and the analytical methods. In Sec.~\ref{sec:results} we present our results for the phase shift caused by dynamical tides, for the comparison with numerical calculations, as well as for the detectability of the phase shifts. We also investigate the measurability of strong first-order phase transitions in the EoS using dynamical tides. Finally, we conclude in Sec.~\ref{sec:summary}.

\section{Tidal model for eccentric binaries}
\label{sec:model}

In the simplest approximation, tidal modes can be treated as forced damped oscillators \cite{Lai:1993di,Flanagan:2007ix}. The dominant source of dissipation in the inspiral phase is expected to be gravitational radiation (see e.g. Refs.~\cite{Flanagan:2007ix,Yang:2018bzx}), which we include in our model. While some of the damping forces caused by bulk or shear viscosities may be important during the merger and post-merger phases \cite{Alford:2019qtm,Alford:2020lla,Most:2022yhe}, they can be neglected during the inspiral phase due to the small viscosities of low-temperature dense matter \cite{Cutler1990,Kochanek:1992wk,Bildsten:1992my}. In the center-of-mass frame dipole moments vanish, while the coupling of external fields to higher-order multipole moments drops quickly. Considering the dominant $l=2$ quadrupole oscillations, the driving force comes from the external quadrupolar tidal field $\mathcal{E}_{ij}$. In this section we introduce our dynamical model by first examining the problem with only classical interactions, then complementing it with general relativistic effects. Throughout the paper we use $G=c=1$.

\subsection{The Newtonian solution}
\label{ssec:Newton}

Let us now consider a binary system with Newtonian interactions, consisting of a NS and another compact object, which can be either a BH or a NS. The components have masses $m_1$ and $m_2$ and Love numbers $\lambda_1$ and $\lambda_2$. For simplicity, we only present the equations for $\lambda_2 = 0$, however, note that our results correspond to binaries, where both NSs have a finite tidal deformability. The tidal Love number and induced quadrupole moment of object $1$ can be expressed as a sum of contributions from $l=2$ modes with various $n\geq0$ radial nodes, $\lambda_1 = \sum_n \lambda_{1,n}$ and $Q_{ij} = \sum_n Q_{ij}^{n}$, where $Q_{ij}$ are the components of the trace-free quadrupole moment tensor, which are related to the density perturbation $Q_{ij}=\int \delta\rho \ (x_i x_j-\frac13 \delta_{ij}) d^3x$ for a Newtonian star. Generally, however, one can neglect contributions from oscillations with $n>0$ radial nodes, since they are subdominant ($\lambda_{1,0}\gg \lambda_{1,n}$ and $\omega_n\geq\omega_0$, for $n>0$) \cite{Flanagan:2007ix}. Hence, we use $\lambda_1 \approx \lambda_{1,0}$ and $Q_{ij} \approx Q_{ij}^0$.

The Lagrangian of the binary system then reads\footnote{see Appendix~\ref{sec:app_grpot} for an explanation of the interaction term}:
\begin{eqnarray}
L &=&  \left[{1 \over 2} \mu {\dot r}^2 + {1 \over 2} \mu
r^2 {\dot
    \varphi}^2 + { m_\mathrm{tot} \mu \over r} \right] -  {1 \over 2} Q_{ij} {\cal
    E}_{ij} \nonumber \\
&& + {1 \over 4 \lambda_{1} \omega_0^2} \left[ {\dot Q}_{ij} {\dot
    Q}_{ij} - \omega_0^2 Q_{ij} Q_{ij} \right],
\label{eq:lagrange}
\end{eqnarray}
where $m_\mathrm{tot}=m_1+m_2$ and $\mu = m_1m_2/m_\mathrm{tot}$ are the total and reduced masses, and $\omega_0$ is the fundamental f-mode frequency. The Lagrangian is expressed in the Newtonian inertial frame, with $r=|\boldsymbol{r}|\equiv|\boldsymbol{x}_1-\boldsymbol{x}_2|$ being the relative distance of the two objects and $\varphi$ is the true longitude, i.e. the angle between $\boldsymbol{e}_x$ and $\boldsymbol{r}$. Note that the static solution for the tidal deformation reads $Q_{ij} = -\lambda \mathcal{E}_{ij}$.

The external tidal field can be expressed as
\begin{eqnarray}
     {\cal E}_{ij} &=&-m_2 \partial_i \partial_j \left( {1 \over r} \right) = -{m_2 \over r^3}(3n^in^j-\delta^{ij}) = \nonumber \\
     &=& -{m_2 \over r^3}\begin{pmatrix}
    {1\over2}+{3\over2}\cos 2\varphi&{3\over2}\sin 2\varphi&0\\
    {3\over2}\sin 2\varphi&{1\over2}-{3\over2}\cos 2\varphi&0\\
    0&0&-1
    \end{pmatrix}.
\end{eqnarray}
Although there are nine elements of the above tensor, it can only have five independent components, due to its symmetrical and traceless nature. Spherical-harmonic tensors, $\boldsymbol{\mathcal{Y}}_{lm}$, give a natural way of representing such objects, since they form a basis for symmetric traceless tensors. They are also related to spherical harmonics through the relation
\begin{equation}
    Y_{lm}(\boldsymbol{n}) = 
\sqrt{\tfrac{(2l+1)!!}{4\pi\, l!}}\
\boldsymbol{\mathcal{Y}}_{lm}\cdot \boldsymbol{n}^{\otimes l} \ ,
\label{Y0}
\end{equation}
where $\boldsymbol{n}$ is a radial unit vector, $\boldsymbol{n}^{\otimes l}=\overbrace{\boldsymbol{n}\otimes \ldots \otimes\boldsymbol{n}}^l$ is the $l$-fold tensor product of $\boldsymbol{n}$, and the inner product is defined for two arbitrary tensors $\boldsymbol{A}$ and $\boldsymbol{B}$ of order $k$ as $\boldsymbol{A}\cdot \boldsymbol{B} = A_{a_1 \ldots a_k}B_{a_1\ldots a_k}$. Another useful property of spherical-harmonic tensors is that they are orthonormal with the inner product:
\begin{equation}
\boldsymbol{\mathcal{Y}}_{lm}\cdot\boldsymbol{\mathcal{Y}}_{lm'}^* = \delta_{mm'} \ ,
\label{Yorth}
\end{equation}
while $\boldsymbol{\mathcal{Y}}_{l(-m)}=(-1)^m\boldsymbol{\mathcal{Y}}_{lm}^*$. One can formulate these tensors directly by considering the Cartesian form of the spherical harmonics, replacing each component $n_x,n_y,n_z$ of the radial unit vector $\boldsymbol{n}$ with Cartesian unit vectors $\boldsymbol{e}_x,\boldsymbol{e}_y,\boldsymbol{e}_z$, the multiplications with tensor products, and normalizing according to Eq.~\eqref{Yorth} (which means factoring in the constant in Eq.~\eqref{Y0}, see Table~\ref{tab:Yexamp} for explicit examples).

\renewcommand{\baselinestretch}{2.0}
\begin{table}[!tb]
\centering
\begin{tabular}[c]{c|c|c}

$lm$ & $Y_{lm}(\boldsymbol{n})$ & $\boldsymbol{\mathcal{Y}}_{lm}$ \\\hline\hline
$2\,0$ & $\sqrt{\frac{5}{4\pi}}(n_z^2-n_\u n_\d)$ & $\sqrt{\frac{2}{3}}(\boldsymbol{e}_z \otimes \boldsymbol{e}_z - \boldsymbol{e}_\u \otimes \boldsymbol{e}_\d)$ \\\hline
$2\,1$ & $-\sqrt{\frac{15}{4\pi}}n_\u n_z$ & $-\sqrt{2} \boldsymbol{e}_\u \otimes \boldsymbol{e}_z$ \\\hline
$2\,2$ & $\sqrt{\frac{15}{8\pi}}n_\u n_\u$ & $\boldsymbol{e}_\u \otimes \boldsymbol{e}_\u$
\end{tabular}
\renewcommand{\baselinestretch}{1}
\caption{\label{tab:Yexamp}Spherical harmonics with Cartesian coordinates and the corresponding spherical harmonic tensors for $l=2$. $n_\u=\frac{1}{\sqrt{2}}(n_x + in_y)$, $n_\d=\frac{1}{\sqrt{2}}(n_x - in_y)$, and similarly $\boldsymbol{e}_\u=\frac{1}{\sqrt{2}}(\boldsymbol{e}_x + i\boldsymbol{e}_y)$, $\boldsymbol{e}_\d=\frac{1}{\sqrt{2}}(\boldsymbol{e}_x - i\boldsymbol{e}_y)$. The normalization factor from Eq.~\eqref{Y0} is $\sqrt{8\pi/15}$. Spherical harmonic tensors for $m=-1,-2$ can be obtained from the identity below Eq.~\eqref{Yorth}.}
\end{table}
\renewcommand{\baselinestretch}{1}

The tensors  $Q_{ij}$ and $\mathcal{E}_{ij}$ are expanded in this basis as
\begin{equation}
\boldsymbol{Q}=\sum_m Q_m \boldsymbol{\mathcal{Y}}_{2m}\,,
\quad
\boldsymbol{\mathcal{E}} = \sum_m \mathcal{E}_m \boldsymbol{\mathcal{Y}}_{2m}\,,\quad
\end{equation}
with coefficients
\begin{equation}
    \mathcal{E}_0 = -\sqrt{3 \over 2}{m_2 \over r^3}\,,\quad
    \mathcal{E}_{\pm2} = -{3 \over 2}{m_2 \over r^3} e^{\mp i2\varphi}\,,\quad
    \mathcal{E}_{\pm1} = 0 \ ,
\end{equation}

From the Lagrangian in Eq.~\eqref{eq:lagrange} the equations of motion for the independent quadrupole components are
\begin{eqnarray}
    {\ddot Q}_m + 2\gamma_0 {\dot Q}_m + \omega_0^2 Q_m &=& -\lambda_1 \omega_0^2 \mathcal{E}_m \ ,
    \label{eq:EoM}
\end{eqnarray}
where we have added a finite but small dissipative term with coefficient $\gamma_0$. We identify this term with gravitational radiation caused by f-mode oscillations. Assuming that $\gamma_0$ is small compared to $\omega_0$ one can calculate $\gamma_0$ using the radiation power for a single oscillation mode
\begin{equation}
    \frac{\mathrm{d}E_\mathrm{GW,Q}}{\mathrm{d}t} = - \frac{1}{5}\left\langle {\dddot Q_{ij} \dddot Q_{ij}} \right\rangle .
\end{equation}
Then for $\gamma_0$ we get
\begin{equation}
    \gamma_0 = \frac15{\lambda \omega_0^6} .
\end{equation}
For realistic NSs, $\gamma_0 \ll \omega_0$ is satisfied.\footnote{In our examples below, we typically assume $\lambda=\Lambda m_{\rm NS}^5$ where $400\leq \Lambda\leq 700$ for which $\gamma_0/\omega_0=\frac15 \Lambda (m_{\rm NS}\omega_0)^5\sim 3\times 10^{-4}$ if $\omega_0=2\pi f_0$ and $1.61\,\mathrm{kHz}\leq f_0\leq 1.81\,\mathrm{kHz}$, see Sec.~\ref{ssec:EoS}.}

One should in principle also account for the spins of the NSs, as they can strengthen dynamical tides \cite{Lai:1993di,Ho:1998hq,Foucart:2018lhe,Ma:2020rak,Steinhoff:2021dsn,Yu:2024uxt}. If the NS is rotating in the retrograde direction with respect to the orbit, the rotation drags the oscillation mode frequencies to lower values in the inertial frame. This allows a resonance to occur at a lower orbital frequency. Studies investigating the effect of spins on dynamical tides have shown that low-frequency g-mode oscillations, which are weakly coupled to the external tides, can be enhanced if the NS rotates faster than $100$~Hz, while f-mode resonance can also significantly increase the phase shift if the spin frequency is higher than $300-500$~Hz, depending on the EoS \cite{Ho:1998hq,Lai:2006pr}. However, most studies, as well as recent observations suggest that a high spin rate is unlikely for NS binaries at the time they enter the aLIGO band, due to spin-down allowed by the long timescale of the binary evolution. For example, both GW events associated with binary NS inspirals, GW170817 and GW190425, were found to be consistent with low spin configurations \cite{LIGOScientific:2018hze,LIGOScientific:2020aai}, while the fastest known spinning pulsar in a binary NS system spins with $44$~Hz \cite{Lyne:2004cj}. In less relativistic systems, such as e.g. white dwarf binaries, external torques acting on dynamical tides can spin up these objects in a way that they become tidally locked to the orbit \cite{Burkart:2012,Burkart:2013,Burkart:2014}. However, for binary NSs the spin-up rate is very low compared to the orbital frequency evolution and tidal locking fails to occur by many orders of magnitude at a wide range of orbital parameters \cite{Burkart:2014,Lai:1993di}. Additionally, the spins of isolated field binaries are expected to be aligned with their orbital angular momentum (e.g. Ref.~\cite{Farr:2017uvj}). Nevertheless, since high eccentricities and misaligned spins are both characteristic of dynamical formation channels, while the inspiral timescale of eccentric binaries are also much shorter than for circular ones, spin effects might still be important for these systems. For the time being, we neglect the effect caused by the rotation of NSs, however, this might be investigated in a follow-up study.

\begin{figure}[t]
    \centering
    \includegraphics[width=0.49\textwidth]{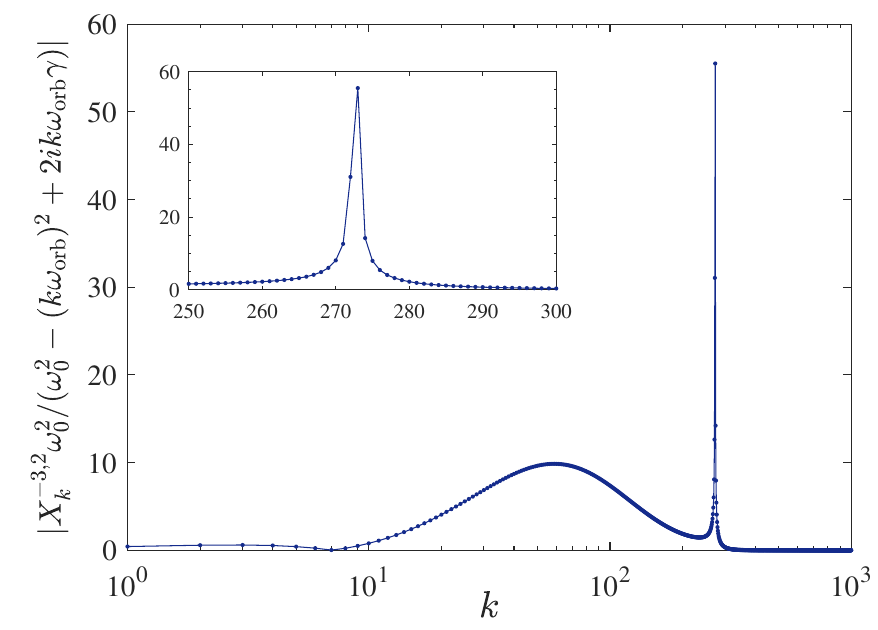}
    \includegraphics[width=0.49\textwidth]{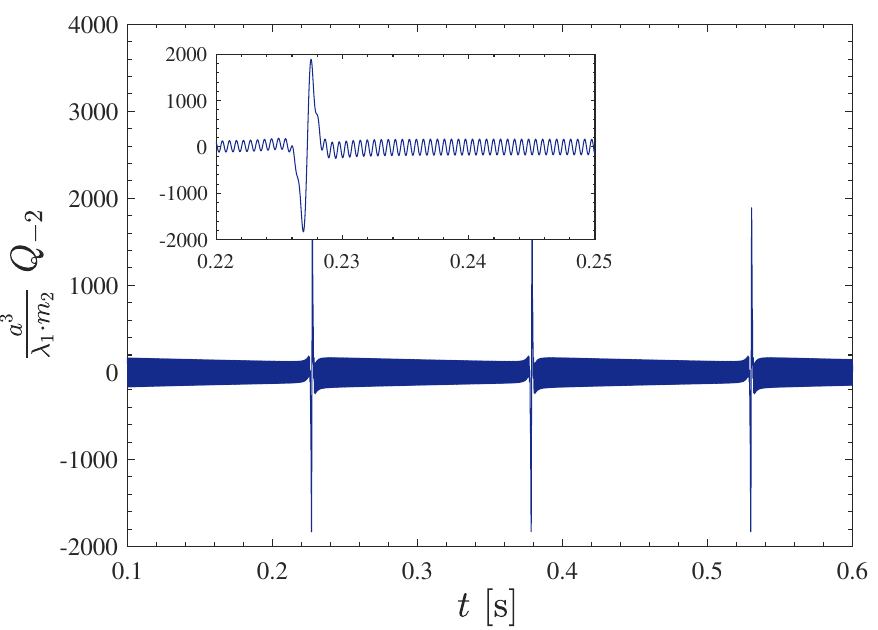}
    \caption{Dimensionless Fourier-coefficients of the quadrupolar deformation $Q_{-2}(t)$ of object $1$, i.e. with respect to the $\boldsymbol{\mathcal{Y}}_{lm}$ mode with $l=2$ and $m=-2$, as a function of the angular frequency $k=\omega/\omega_\mathrm{orb}$ (top) and the time dependence of $Q_{-2}(t)$ (bottom) in a binary NS system with $m_1=m_2=1.4~\Msun$. The orbital parameters are fixed at $a = 600$~km, $e=0.9$, while the tidal parameters are $\Lambda_1=\lambda_1/m_1^5=400$ and $f_0=1800$~Hz (based on Ref.~\cite{Pradhan:2022vdf}). The orbital angular frequency is then $\omega_\mathrm{orb}\approx41.5~s^{-1}$, which means that the resonance peak is at $k_\mathrm{res}=2\pi f_0/\omega_\mathrm{orb}\approx272.7$. The values of the discrete Fourier spectrum are connected to guide the eye.}
    \label{fig:Hansen}
\end{figure}

As a first approximation in describing dynamical tides, we neglect the back-reaction effect of the tidal deformations on the orbital parameters. Then the orbital motion will be periodic and hence can be expanded as a Fourier-series in time, or equivalently, in the mean anomaly, $M$.\footnote{To avoid confusion note that we are using $m_{1,2}$ and $m_{\rm tot}=m_1+m_2$ to describe mass parameters; $M$ is used exclusively to denote mean anomaly.} For this purpose, we utilize the so-called Hansen-coefficients, $X_k^{q,m}$, which are defined as:
\begin{equation}
    \left( \frac{r}{a} \right)^q \exp(im\varphi) = \sum\limits_{k=-\infty}^{\infty} X_k^{q,m}(e) \ \mathrm{exp}(ikM) \ ,
\end{equation}
where $q$ is an arbitrary integer and $e$ is the eccentricity of the orbit. These coefficients are calculated using the following relation (assuming $M=0$ at periapsis):
\begin{equation}
    X_k^{q,m}(e) = \frac{1}{2\pi} \int\limits_{-\pi}^{\pi} \left( \frac{r}{a} \right)^q \cos(m\varphi - kM) \mathrm{d}M \ ,
\end{equation}
Therefore, the right hand side of Eq.~\eqref{eq:EoM} can be expressed as
\begin{equation}
    -\lambda_1 \omega_0^2 \mathcal{E}_m(t) = C_m \lambda_1 \omega_0^2 \frac{m_2}{a^3} \sum\limits_{k=-\infty}^{\infty} X_k^{-3,-m} e^{ikM} \ ,
    \label{eq:E_X}
\end{equation}
where $C_m = 3/2$ for $m=\pm2$ and $C_m=\sqrt{3/2}$ for $m=0$. Hence, the time domain solutions for the tidal components are:
\begin{align}
    Q_m(t) &= C_m \lambda_1 \omega_0^2 \frac{m_2}{a^3} \times \nonumber \\
    &\times \sum\limits_{k=-\infty}^{\infty} \frac{X_k^{-3,-m}}{\omega_0^2 - (k\omega_\mathrm{orb})^2 + 2i\gamma_0 k\omega_\mathrm{orb}} e^{ikM} \ ,
    \label{eq:Qanalytic}
\end{align}
where $\omega_\mathrm{orb}=\sqrt{m_\mathrm{tot}/a^3}$ is the orbital angular frequency. Note, that the implicit time dependency is in $M\equiv M(t)=\omega_\mathrm{orb}t$. Practically, due to the exponential suppression of the spectrum at large values of $k$ (see Fig.~\ref{fig:Hansen}), we only need to sum a finite number of contributions to obtain $Q(t)$ up to a given precision.

Fig.~\ref{fig:Hansen} shows an example for the Fourier components of $Q_{-2}$, while the time dependence of $Q_{-2}$, obtained from Eq.~\eqref{eq:Qanalytic} is also shown in the bottom panel. The Fourier components exhibit a broad plateau due to the high eccentricity, while there is also a narrow peak corresponding to $f_0=\omega_0/2\pi$. We can also see the necessity of an eccentric orbit for the excitation of dynamical tides. Without an eccentric orbit, the spectrum of the tidal field would consist of a single peak (at $-mf_{\text{orb}}$), hence, modes with a frequency of $f_0$ would not be excited unless the system is close to resonance ($f_{\text{orb}} = f_0/2$ for $|m|=2$). However, since $f_0\gg f_\mathrm{orb}$ during most of the inspiral of binary NSs, the system would not reach a resonant state before approaching its merger. From the bottom panel of Fig.~\ref{fig:Hansen} one can easily understand what is happening during eccentric encounters. The compact objects spend most of their orbit far from each other, hence, tidal forces are weak for most of the orbit. When the NSs approach pericenter, tidal forces suddenly increase and then decrease immediately. Therefore, the neutron stars act as bells which are struck once during each orbit. Tidal modes get a short impulse during the pericenter passage and then oscillate freely for the rest of the orbit, while also slowly losing energy due to GW radiation.

\subsection{General relativistic effects}
\label{ssec:GRan}

In addition to nonrelativistic tidal forces, general relativistic effects are also relevant during this process, such as energy loss due to gravitational radiation. The orbit of an eccentric binary system with semi-major axis $a$ and eccentricity $e$ and without tidal effects is modified due to gravitational radiation  according to the following equations in the 2.5 Post-Newtonian (PN) leading order \cite{Peters:1964zz}:
\begin{equation}
\frac{\mathrm{d}a}{\mathrm{d}t} = - \frac{64}{5} \frac{m_1 m_2 m_\mathrm{tot}}{a^3 (1-e^2)^{7/2}} \left( 1+ \frac{73}{24} e^2+ \frac{37}{96} e^4 \right) ,
\label{eq:aevolution}
\end{equation}
\begin{equation}
\frac{\mathrm{d}e}{\mathrm{d}t} = - \frac{304}{15} \frac{m_1 m_2 m_\mathrm{tot}e}{a^4 (1-e^2)^{5/2}} \left( 1+ \frac{121}{304} e^2 \right) .
\label{eq:eevolution}
\end{equation}
It is worth introducing the dimensionless pericenter distance $\rho_\mathrm{p} = a(1-e)/m_\mathrm{tot}$. Eqs. \eqref{eq:aevolution} and \eqref{eq:eevolution} can be solved analytically and yield the following relation for $\rho_\mathrm{p}(e)$ \cite{Peters:1964zz,OLeary:2008myb}:
\begin{equation}
\rho_{\mathrm{p}}(e) = \frac{c_0 e^{12/19}}{(1+e)} \left( 1+\frac{121}{304}e^2 \right)^{\frac{870}{2299}} ,
\label{eq:rhoe}
\end{equation}
where $c_0$ is a dimensionless constant that can be expressed from the initial parameters, setting $e=e_0$ and $\rho_{\mathrm{p}}=\rho_{\mathrm{p0}}$ in Eq.~(\ref{eq:rhoe}). 
                        
The inspiral ends either at physical collision or when reaching the last stable orbit (LSO) determined by\cite{Cutler:1994pb}:
\begin{equation}
    \rho_\mathrm{p}(e_\mathrm{LSO}) = \frac{6 + 2 e_\mathrm{LSO}}{1 + e_\mathrm{LSO}}.
    \label{eq:LSO}
\end{equation}

The orbit-averaged power radiated through GWs is
\begin{equation}
    \frac{\mathrm{d}E_{\mathrm{GW}}}{\mathrm{d}t} = - \frac{32}{5} \frac{\mu^2 m_\mathrm{tot}^3}{a^5(1-e^2)^{7/2}} \left( 1+\frac{73}{24}e^2 +\frac{37}{96}e^4 \right),
    \label{eq:dE_GW}
\end{equation}
therefore, the energy loss during one orbit is $\Delta E_\mathrm{GW} = \dot{E}_\mathrm{GW}\cdot2\pi/\omega_\mathrm{orb}$.

These formulas assume the leading order quadrupole formula for radiation-reaction where the objects follow Newtonian trajectories with slowly decaying $a$ and $e$. We examine the impact of other terms in a post-Newtonian (PN) expansion in powers of $v/c$ which further modify the elliptical orbit and examine the corresponding GW phase shift in Sec.~\ref{ssec:numsim} (see also \cite{Lincoln:1990ji,Vick:2019cun}). In particular, we account for the conservative 1PN and 2PN general relativistic corrections to the orbital motion and the dissipative $2.5$PN GW radiation back-reaction terms, and also include the backreaction of tidally excited $f$-mode stellar oscillations in our numerical simulations, given explicitly in Appendix~\ref{sec:appendix}. 

Note that for a proper treatment of the post-Newtonian interactions we would need to account for additional effects. First of all, in order to correctly determine the radiation back-reaction terms, we would need to use the derivatives of the total quadrupole moment instead of calculating the back-reaction terms separately for the orbit and the tides (see e.g. Refs.~\cite{Flanagan:2007ix,Blanchet:2013haa}). This would induce cross-terms between the orbital and tidal evolution. The impact of the post-Newtonian potential on the tides is another possibly important effect to take into account \cite{Vines:2010ca,Vines:2011ud,Henry:2019xhg}. In this paper we neglect these terms and concentrate on the leading-order evolution, however, these effects could be taken into consideration in a follow-up analysis.

\subsection{Modeling NS properties}
\label{ssec:EoS}

Both the tidal deformability $\lambda$ and the f-mode frequency $f_0$ are dependent on the EoS describing the matter inside NSs. However, various quasi-universal relations can be drawn between different macroscopic properties of NSs \cite{Yagi:2013awa}. The tidal deformability of $m_\mathrm{NS}=1.4$~$\Msun$ NSs has been recently constrained to $\Lambda(1.4~\Msun)<720$ by the observation of GW170817 \cite{LIGOScientific:2018hze}, where $\Lambda = \lambda/m_\mathrm{NS}^5$. We expect the $f_0$ frequency of $1.4$~$\Msun$ NSs to fall between 1.5~kHz and 2.5~kHz \cite{Pradhan:2022vdf}. In this paper, instead of calculating tidal deformabilities and f-mode frequencies individually for specific equations of state, we utilize quasi-universal relations between various properties of NSs. We also examine possible deviations from these relations due to first-order phase transitions or quark star solutions.

Most of our calculations are done using binary NSs with two $1.4~\Msun$ components. In this case we only need to find the connection between $\Lambda(1.4~\Msun)$ and $f_0(1.4~\Msun)$. This is done by utilizing the quasi-universal relation derived in Ref.~\cite{Pradhan:2022vdf} for NSs containing nuclear or hyperonic matter. This relation gives us much more accurate results than using general relations between these and other properties of NSs.

However, in order to be able to calculate f-mode frequencies and tidal deformabilities for NSs with arbitrary masses, we need to apply a two-step process. After specifying the mass and the radius of the NS, first we calculate the tidal deformability from the compactness, $C\equiv M/R$, inverting the quasi-universal relation presented in Ref.~\cite{Godzieba:2020bbz}. Then, we calculate the f-mode frequency as well, utilizing the relation derived in Ref.~\cite{Pradhan:2023zmg} between $C$ and $M\omega_0$. Other parameterizations also exist for these quasi-universal relations (see \textit{e.g.} Refs.~\cite{Pradhan:2022vdf,Yagi:2016bkt,Maselli:2013mva,Urbanec:2013fs}), however, we chose these ones due to their generality in the choice of constructing the EoS, allowing first-order phase transitions as well. Both of these relations apply within an error of $\sim5-10\%$. Crosschecking this two-step process using the quasi-universal relation between $\Lambda(1.4~\Msun)$ and $f_0(1.4~\Msun)$, we find that it overestimates the f-mode frequency by $\sim5\%$. Since the values of $\Lambda(1.4~\Msun)$ and $f_0(1.4~\Msun)$ show a tighter correlation, we assume that $f_0(1.4~\Msun)$ can be estimated more accurately from the quasi-universal relation of Ref.~\cite{Pradhan:2022vdf}.

\section{Results}
\label{sec:results}

Dynamical tides excited by the eccentric orbital motion may be directly observed through their GW emission. However, as we will see later, these GW signals alone are typically below the detection threshold of current GW detectors. Another way to observe these dynamical tides with GW instruments is indirectly through their effect on the orbital evolution. While the amplitude of the GW signals emitted by the binary system is not expected to change much, the effect of these dynamical tides on the phase shift may be much more pronounced. 

In this section we provide analytical formulas for the energy transferred to the tidal modes and the phase shift caused by this energy transfer. We also compare the analytical results with our numerical simulation, calculating the signal-to-noise ratio (SNR) of the phase shift, and examine the possible observational implications for the EoS of NSs.

First we construct simple leading-order typically Newtonian analytic models to interpret the results of the numerical simulations. We investigate post-Newtonian effects in Sec.~\ref{ssec:numsim}.

\subsection{Energy transmitted to tidal modes}

To calculate the energy transferred from the orbital motion to tidal modes, let us consider the binary system again without GR effects and energy loss due to GW radiation. Then the change in the energy of a single oscillation mode is
\begin{align}
    \Delta E &= \Delta \left[ \frac{1}{4\lambda_1\omega_0^2} \left( \dot{Q}_{ij}\dot{Q}_{ij} + \omega_0^2 Q_{ij} Q_{ij} \right) \right] = \nonumber \\ 
    &= \frac{1}{2\lambda_1\omega_0^2} \int\limits_{t_1}^{t_2}\mathrm{d}t \ \dot{Q}_{ij} \left( \ddot{Q}_{ij} + \omega_0^2 Q_{ij} \right) = \nonumber \\
    &= -\frac{1}{2} \int\limits_{t_1}^{t_2}\mathrm{d}t \ \mathcal{E}_{ij} \dot{Q}_{ij} \ .
\end{align}
Following \cite{Press1977}, let $t_1\rightarrow-\infty$ and $t_2\rightarrow\infty$. Then, using the Fourier-transforms of $\mathcal{E}$ and $Q$ defined as
\begin{equation}
    f(t) = \int\limits_{-\infty}^\infty \mathrm{d}\omega \ e^{-i\omega t} \Tilde{f}(\omega) \ ,
\end{equation}
we get \cite{Press1977}:
\begin{align}
    &\Delta E = \pi\lambda_1\omega_0^2 \int\limits_{-\infty}^\infty \mathrm{d}\omega \ \frac{-i\omega |\Tilde{\mathcal{E}}_{ij}(\omega)|^2}{\omega_0^2 - \omega^2 - i\omega\epsilon} = \nonumber \\
    &= \pi^2\lambda_1\omega_0^2 |\Tilde{\mathcal{E}}_{ij}(\omega_0)|^2 = \sum_m \pi^2\lambda_1\omega_0^2 |\Tilde{\mathcal{E}}_m(\omega_0)|^2 \ ,
    \label{eq:DeltaE}
\end{align}
where in the periodic case from Eq.~\eqref{eq:E_X},
\begin{equation}
    \Tilde{\mathcal{E}}_m(\omega) = - \sum\limits_{k=-\infty}^{\infty}C_m \frac{m_2}{a^3} X_{-k}^{-3,-m} \delta(\omega - k\omega_\mathrm{orb}) \ .
\end{equation}
Hence, in order to get the energy transferred to tidal modes, one needs to evaluate the Fourier-transform of the tidal field at the resonance frequency. Note that in case we have a periodic excitation where none of the harmonics coincide with the resonance frequency, the total energy transmitted is zero. In contrast, the transmitted energy becomes infinity as these two coincide, which corresponds to a resonant excitation.

However, we are interested in the energy exchange during a single passage. Then, the Fourier transform should be limited to only a single orbital period, while keeping the tidal field zero elsewhere. However, in many cases, especially close to the merger, where the tidal field at the apocenter is not negligible, the jump at the boundary results in an oscillatory spectrum, and therefore the energy exchange will also include this artificial oscillation. In fact, it is not trivial how to uniquely define the single-passage scenario in the moderate and low-eccentricity case, where tidal forces in the apocenter are non-negligible, without considering the entire evolution of the system. We find that a convenient way to define the energy transfer is to smoothly interpolate between the harmonics of the discrete Fourier spectrum. This way, we get a continuous and smooth spectrum, while this smooth interpolation also creates a natural smoothing at the edges of the tidal field in the time domain. We can then obtain the coefficient in Eq.~\eqref{eq:DeltaE} for all values of $\omega_0$ (for which $\omega_0 = k\omega_{\rm orb}$ does not necessarily hold) by interpolating the discrete function:
\begin{equation}
    \tilde{\mathcal{E}}_m(\omega = k\omega_\mathrm{orb}) = - C_m \frac{m_2}{\omega_\mathrm{orb} a^3} X_{-k}^{-3,-m} \: ,
    \label{eq:DeltaE2}
\end{equation}
where the extra factor $1/\omega_\mathrm{orb}$ comes from smearing the $\delta$ functions separated from each other by an $\omega_\mathrm{orb}$ wide interval. The change in the orbital energy due to tidal excitations is then:
\begin{equation}
    \Delta E_Q = -\Delta E \: .
    \label{eq:DeltaEQ}
\end{equation}
Since the tidal oscillations of the two NSs are independent of (not coupled to) each other, when taking into account the quadrupole deformation of both objects, the energies extracted from the orbit simply add up. Comparing to our numerical simulations, we find that this formula remains accurate (with a relative error of $\lesssim10\%$) even for eccentricities as low as $e\approx0.2$.

\begin{figure}[t]
    \centering
    \includegraphics[width=0.49\textwidth]{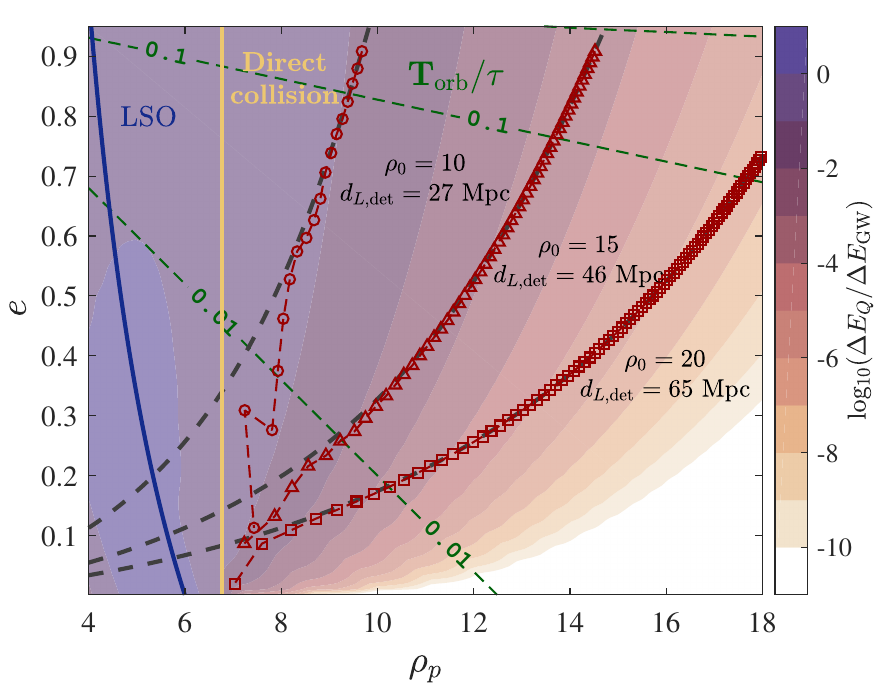}
    \caption{The logarithm of $\Delta E_Q/\Delta E_\mathrm{GW}$ as a function of orbital parameters $\rho_\mathrm{p}$ and $e$ for two $1.4$~$\Msun$ NSs. The dashed black lines show the orbit-averaged evolution of binaries with selected initial conditions $e_0\approx 1$ and $\rho_\mathrm{p0}=(10,15,20)$, driven by Eqs.~\ref{eq:aevolution}-\ref{eq:eevolution}. The dashed red lines with different markers correspond to compact binaries with the same initial parameters but with tidal interactions also taken into account. These come from our numerical simulation, and each marker corresponds to an apocenter passage. The tidal deformability and the f-mode frequency was chosen as $\Lambda=400$ and $f_0=\omega_0/2\pi=1800$~Hz. The blue line represents parameters at the last stable orbit, while the vertical yellow line corresponds to direct collision of the two NSs, where we used the very conservative estimate $R_\mathrm{NS}(1.4~\Msun)=14$~km. Dashed green lines show different values of $T_\mathrm{orb}/\tau$. For each selected orbital evolution we display the sky position and orientation averaged detection range with a single aLIGO detector, $d_{\rm L, det}$, calculated from the orbit-averaged solution \cite{OLeary:2008myb}.}
    \label{fig:dE}
\end{figure}

\begin{figure}[t]
    \centering
    \includegraphics[width=0.49\textwidth]{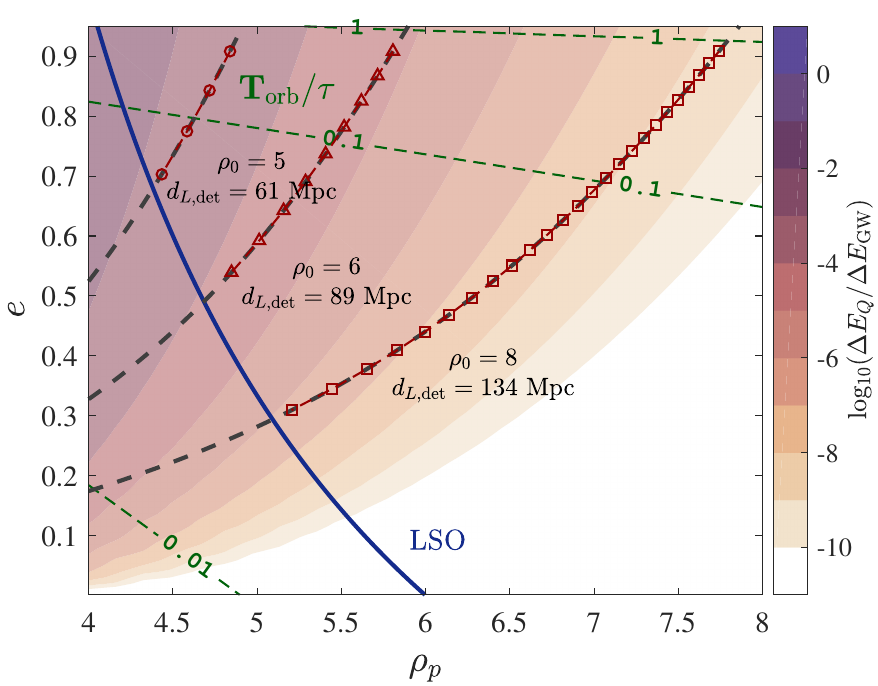}
    \caption{Same as Fig.~\ref{fig:dE} but for a binary consisting of a NS with $m_1 = 1.4~\Msun$ and a BH with $m_2 = 10~\Msun$. The three selected orbital evolutions here have $e_0\approx1$ and $\rho_\mathrm{p0}=(5,6,8)$. The tidal parameters of the NS are $\Lambda=400$ and $f_0=1800$~Hz.}
    \label{fig:dEBH}
\end{figure}

To understand the amount of energy transmitted to tidal modes during a single orbit, $\Delta E_Q$, it is useful to compare it to the amount of energy radiated away during a single orbit, $\Delta E_\mathrm{GW}=\dot E_\mathrm{GW} T_{\rm orb}$ (Eq.~\ref{eq:dE_GW}) where  $T_\mathrm{orb}=2\pi/\omega_\mathrm{orb}$ is the orbital period. The color maps in Figs.~\ref{fig:dE} and \ref{fig:dEBH} show the logarithm of $\Delta E_Q/\Delta E_\mathrm{GW}$ as a function of $\rho_\mathrm{p}$ and $e$ for a binary of two $1.4~\Msun$ NSs and one with such a NS and a $10~\Msun$ BH, respectively.

One can observe that there is an exponential-like dependency of $\Delta E_Q/\Delta E_\mathrm{GW}$ as a function of $\rho_\mathrm{p}$. This can be attributed to the fact that the Hansen coefficients have an exponential tail for large values of $k$, and increasing $\rho_\mathrm{p}$ means increasing $\omega_0/\omega_\mathrm{orb}$ as well. Looking at the numerical simulation results, the tidal interaction becomes visible above $\Delta E_Q/\Delta E_\mathrm{GW}\gtrsim10^{-3}$. Another observation can be made by looking at the values of $T_\mathrm{orb}/\tau$, where $\tau=1/\gamma_0$ is the damping time. For most of the parameter space $T_\mathrm{orb}/\tau\ll 1$, which means that tidal oscillations do not have time to attenuate from one pericenter passage to the next. However, the oscillation may also be cancelled by the subsequent pulse-like excitation if it comes at the appropriate phase. Due to the continuously changing orbital period, the phase at which the tidal pulse arrives is seemingly random. Thus, tidal excitations will appear to follow a random walk and so will the energy extracted from the orbit. Similar tidal evolutions were found in Refs.~\cite{Vick:2019cun,Samsing:2016bqm}. The red lines with markers in Fig.~\ref{fig:dE} show orbital evolutions when tidal effects are taken into account which deviate from simulations without tidal effects (dashed black lines) for small $\rho_p$.  Note that this figure corresponds to NSs with tidal parameters $\Lambda=400$ and $f_0=1800$~Hz, and the excitation energy would be increased if $\Lambda$ was higher or $f_0$ lower.

Fig.~\ref{fig:dEBH} shows the same for binaries consisting of a $1.4~\Msun$ NS and a $10~\Msun$ BH. The message of this figure is clear. Dynamical tides play a much less prominent role for NS-BH binaries. The reason is twofold. There is only one NS, which can go through substantial tidal oscillations, and the orbital frequency of the system for given orbital parameters become lower, which, due to the exponential dependence of the Hansen coefficients on $k$ results in much lower excitation energies. Fig.~\ref{fig:dEBH} also shows that NSs are not expected to be tidally disrupted in a NS--BH binary unless the BH is highly spinning and the EoS of matter inside the NS is stiff at low densities -- which would result in a larger radius and $\Lambda$, and a lower f-mode frequency.


\subsection{Phase shift caused by tidal interactions}

Since the energy transmitted to tidal modes remain low during most of the orbital evolution and can only exceed $\Delta E_\mathrm{GW}/10$ during the last couple of orbits, and since the frequency of the tidal modes falls in the kHz band where aLIGO is less sensitive, GWs produced by these oscillations have a low SNR. However, due to back-reaction, the orbital evolution can be modified substantially. GW detectors are particularly sensitive to changes in the GW phase. We derive analytical formulas for the phase shift caused by the back-reaction from tidal excitations and examine their detectability by GW detectors. A common rule of thumb is to check if the phase shift exceeds 1 radian at any time during the detectable part of the evolution. If it does, and the binary is detectable, then the phenomenon causing the phase shift is expected to be detectable.

A common way to quantify the phase shift is to determine its dependence on the orbital frequency \cite{Flanagan:2007ix,Kocsis:2011dr,Chirenti:2016xys,Yang:2018bzx,Yang:2019kmf}. Some other studies investigated the effect of g-mode excitations in eccentric binaries or resonant g-mode excitations in circular ones \cite{Lai:1993di,Xu:2017hqo,Andersson:2017iav,Kuan:2021jmk}. Here we follow a similar method. Let us consider a binary system whose orbital energy is decreasing due to GW emission and some energy is transferred to tidal modes. Then, the change in the mean anomaly during an infinitesimally short period of time is
\begin{align}
    \mathrm{d} M &= \omega_\mathrm{orb} \: \mathrm{d}t = \omega_\mathrm{orb} \frac{\mathrm{d}E_\mathrm{orb}}{\dot{E}_\mathrm{orb}} = \omega_\mathrm{orb} \frac{\mathrm{d}E_\mathrm{orb}}{\dot{E}_\mathrm{GW}+\dot{E}_Q} \approx \nonumber \\
    &\approx \omega_\mathrm{orb} \frac{\mathrm{d}E_\mathrm{orb}}{\dot{E}_\mathrm{GW}} \left( 1-\frac{\dot{E}_Q}{\dot{E}_\mathrm{GW}} \right)\: .
    \label{eq:dM}
\end{align}
The phase difference between systems with and without tidal excitations is:
\begin{equation}
    \mathrm{d} (\delta M) \approx - \omega_\mathrm{orb} \frac{\mathrm{d}E_\mathrm{orb}}{\dot{E}_\mathrm{GW}} \frac{\dot{E}_Q}{\dot{E}_\mathrm{GW}} \approx - \omega_\mathrm{orb} \frac{\mathrm{d}E_Q}{\dot{E}_\mathrm{GW}} \: ,
    \label{eq:ddM}
\end{equation}
where $\mathrm{d}E_Q$ is the change in the energy due to tidal oscillations while the orbital energy changes by $\mathrm{d}E_\mathrm{orb}$. Note that the integration variable is $E_\mathrm{orb}$ in both cases, hence, this formula applies for orbital phases compared at the same orbital energies, or equivalently, the same orbital frequency. This will become important later, when comparing this formula to numerical results. Also note that since both $\mathrm{d}E_Q$ and $\dot{E}_\mathrm{GW}$ are negative, the phase difference will be negative as well, i.e. with dynamical tides it takes fewer orbits to get to a certain orbital frequency.

During a single orbit, the phase shift then becomes
\begin{equation}
    \delta M_1 \approx -\omega_\mathrm{orb} \frac{\Delta E_Q}{\dot{E}_\mathrm{GW}} = -2\pi \frac{\Delta E_Q}{\Delta E_\mathrm{GW}} \: ,
    \label{eq:deltaM1}
\end{equation}
where $\Delta E_Q/\Delta E_\mathrm{GW}$ is the quantity derived in the previous section.

Now we also need to take into account that while the evolution of the tides is fundamentally deterministic, due to the seemingly random phase at which the short tidal pulses arrive, the amplitude of tidal excitations will appear to vary stochastically. The $j^\mathrm{th}$ pulse can be characterized by the complex number $\Delta q_j = A_j \mathrm{exp}(i\varphi_j)$, where $\varphi_i$ is a random phase in the $[0,2\pi]$ interval. Thus, the expectation value of the complex amplitude after the $j^\mathrm{th}$ pulse is
\begin{equation}
    \langle q_j \rangle = \sum\limits_{k=1}^j \langle \Delta q_k \rangle = 0 .
\end{equation}
Since the energy of tidal modes is quadratic in the amplitude, its expectation value over the incoherent random realizations of the tidal pulses is
\begin{align}
    \langle E_{Q,j} \rangle &= C \left\langle \left|\sum\limits_{k=1}^j \Delta q_k \right|^2 \right\rangle = C \sum\limits_{k=1}^j \left\langle \left| \Delta q_k \right|^2 \right\rangle = \nonumber \\
    &= C \sum\limits_{k=1}^j A_k^2 = \sum\limits_{k=1}^j \Delta E_{Q,k} \: ,
\end{align}
where we have used the fact that the expectation value of cross terms gives zero. $\Delta E_{Q,k}$ is the energy transmitted to tidal modes with zero initial amplitude. Utilizing Eq.~\eqref{eq:deltaM1},
\begin{equation}
    \langle \delta M \rangle = \sum\limits_{k=1}^j \delta M_{1,k} \approx \sum\limits_{k=1}^j \left( -2\pi \frac{\Delta E_Q}{\Delta E_\mathrm{GW}} \right)_k \: .
    \label{eq:deltaM_an}
\end{equation}
Hence, to get the expectation value of the phase shift, we must add up the phase shift contributions of each orbit. Note that this result would be the same if the tidal modes were strongly attenuated.

Let us also utilize the orbit-averaged evolution of the binary without tidal interactions in Eq.~\eqref{eq:dM} to get an approximate integral formula as
\begin{align}
    \mathrm{d} (\delta M) &=
    - \omega_\mathrm{orb} \frac{dE_{\rm orb}}{df_{\rm orb}} df_{\rm orb}\frac{\dot{E}_Q}{\dot{E}_\mathrm{GW}^2}
    + \mathcal{O}\left(\frac{\dot{E}_Q^2}{\dot{E}_\mathrm{GW}^3}\right)\nonumber\\
    &= \frac{2\pi}3 \mathcal{M}^{5/3}\omega_{\rm orb}^{2/3} df_{\rm orb}\frac{\dot{E}_Q}{\dot{E}_\mathrm{GW}^2}+ \mathcal{O}\left(\frac{\dot{E}_Q^2}{\dot{E}_\mathrm{GW}^3}\right)
    \: ,
\end{align}
where we used the leading order (Newtonian) orbital energy $E_{\rm orb}=-\frac12 \eta m_{\rm tot}^2/a=-\frac12 \eta m_{\rm tot}^{5/3}\omega_{\rm orb}^{2/3}=-\frac12 \mathcal{M}^{5/3}\omega_{\rm orb}^{2/3}$ and Kepler's law $a =  m_{\rm tot}^{1/3}\omega^{-2/3}$, hence $\omega_{\rm orb} dE_{\rm orb}/df_{\rm orb} = 2\pi \omega_{\rm orb} dE_{\rm orb}/d\omega_{\rm orb} = 2\pi \frac23 E_{\rm orb} %= -2\pi \frac23 \frac12  \mathcal{M}^{5/3}\omega_{\rm orb}^{2/3} = 
=-\frac23 \pi \mathcal{M}^{5/3}\omega_{\rm orb}^{2/3}$. Thus:
\begin{equation}
    \langle \delta M \rangle (f)\approx \frac{(2\pi)^{5/3}}{3}\int\limits_{f_\text{in}}^{f} \mathrm{d}f_\mathrm{orb} \: 
     \mathcal{M}^{5/3} f_\mathrm{orb}^{2/3}  \frac{\dot{E}_Q}{\dot{E}_\mathrm{GW}^2} \: ,
    \label{eq:deltaM_orbave}
\end{equation}
where $\mathcal{M}=m_{\rm tot}^{2/5}\mu^{3/5}$ is the chirp mass, $\dot{E}_\mathrm{GW}$ is given by Eq.~\eqref{eq:dE_GW} and $\dot{E}_Q = \Delta E_Q / T_\mathrm{orb}$, where $\Delta E_Q$ is given by Eqs.~\eqref{eq:DeltaE}, \eqref{eq:DeltaE2}, and \eqref{eq:DeltaEQ}. Note that these expressions only account for the phase shift caused by dynamical tides and do not include adiabatic tides. The effect of adiabatic tide can be calculated separately, and has been done by Ref.~\cite{Flanagan:2007ix} for circular and by Ref.~\cite{Bernaldez:2023xoh} for eccentric binaries.

\begin{figure}[t]
    \centering
    \includegraphics[width=0.49\textwidth]{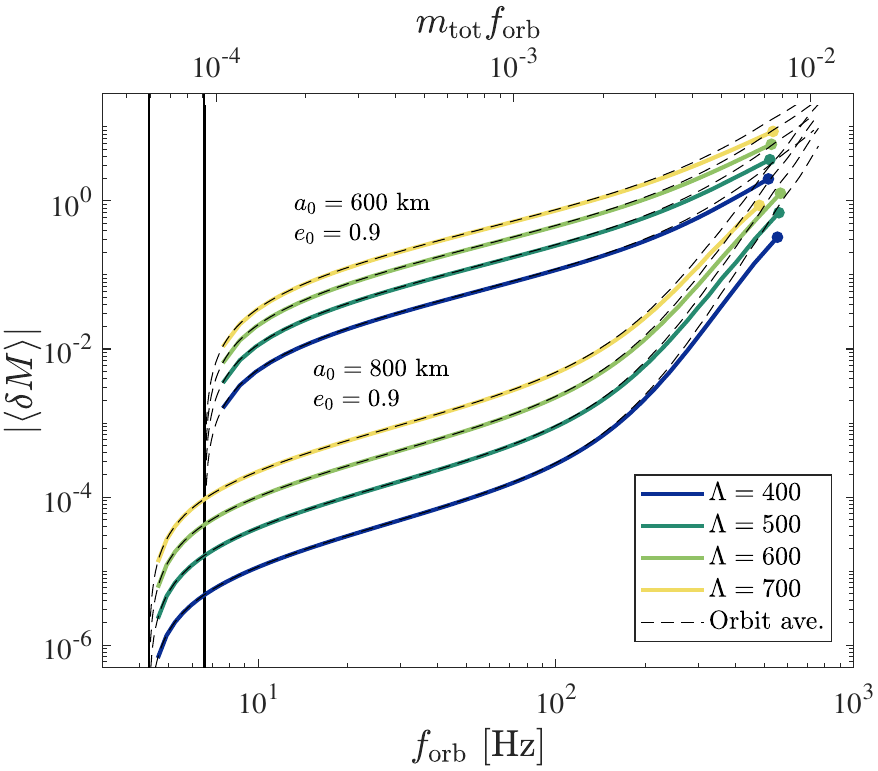}
    \caption{The mean phase shift $\langle\delta M\rangle$ caused by tidal oscillations as a function of the orbital frequency for binary NSs with two $1.4~\Msun$ components, calculated using Eq.~\eqref{eq:deltaM_an} with the input orbital parameters ($a$ and $e$, see Appendix~\ref{sec:appendix}) extracted from our numerical simulations run in the adiabatic tide limit, i.e. with $f_0 = f_\mathrm{ad} \gg f_\mathrm{orb}$ (solid lines) and the orbit-averaged analytical calculation Eq.~\eqref{eq:deltaM_orbave} (dashed lines). Different colors correspond to different equations of state, and hence different values for $\Lambda$ and $f_0$. For $\Lambda=(400,500,600,700)$ the corresponding values for $f_0$ are $1.81$~kHz, $1.73$~kHz, $1.66$~kHz and $1.61$~kHz, respectively. The filled circles denote the last apocenter passage before the two NSs collide directly, which we defined as their distance dropping below $28$~km. The black vertical lines denote the initial orbital frequencies for the two different orbital evolutions.}
    \label{fig:deltaM_an}
\end{figure}

Fig.~\ref{fig:deltaM_an} shows the orbital phase shift as a function of orbital frequency for binary NSs with two $1.4~\Msun$ components. The figure shows the results for several different $\Lambda-f_0$ pairs and two different orbital configurations. The solid lines denote results obtained from Eq.~\eqref{eq:deltaM_an}, where the input orbital parameters (i.e. $a$ and $e$ to be used for calculating $\dot{E}_Q$ and $\dot{E}_\mathrm{GW}$) were extracted from our numerical simulations in the adiabatic tide limit, i.e. with $f_0 = f_\mathrm{ad} \gg f_\mathrm{orb}$\footnote{Since $f_\mathrm{orb}^\mathrm{max}\leq 800$~Hz in our configurations, we find that the adiabatic limit is already approached for $f_\mathrm{ad}\geq2.5$~kHz, while keeping the integration time reasonably low.}, while the dashed lines denote results obtained with Eq.~\eqref{eq:deltaM_orbave}, for which the orbit-averaged evolution was used without taking into account the effect of adiabatic tides. There is a good agreement between the two formulae at moderately high orbital frequencies, however, at the highest frequencies the orbit-averaged formula overestimates the results calculated from our numerical simulations. This is due to the effect of adiabatic tides, which influence the orbital evolution even in the $f_0\gg f_\mathrm{orb}$ limit. It is clearly visible that increasing the tidal deformability has a significant effect on the magnitude of the phase shift. Changing the tidal deformability from $\Lambda=400$ to $700$ increases the phase shift by a factor of $\sim5$ near the merger of the binary. This is due to the decrease of the f-mode frequency with increasing $\Lambda$, which has a significant effect on the amount of energy transmitted to tidal oscillation modes.

\begin{figure}[t]
    \centering
    \includegraphics[width=0.49\textwidth]{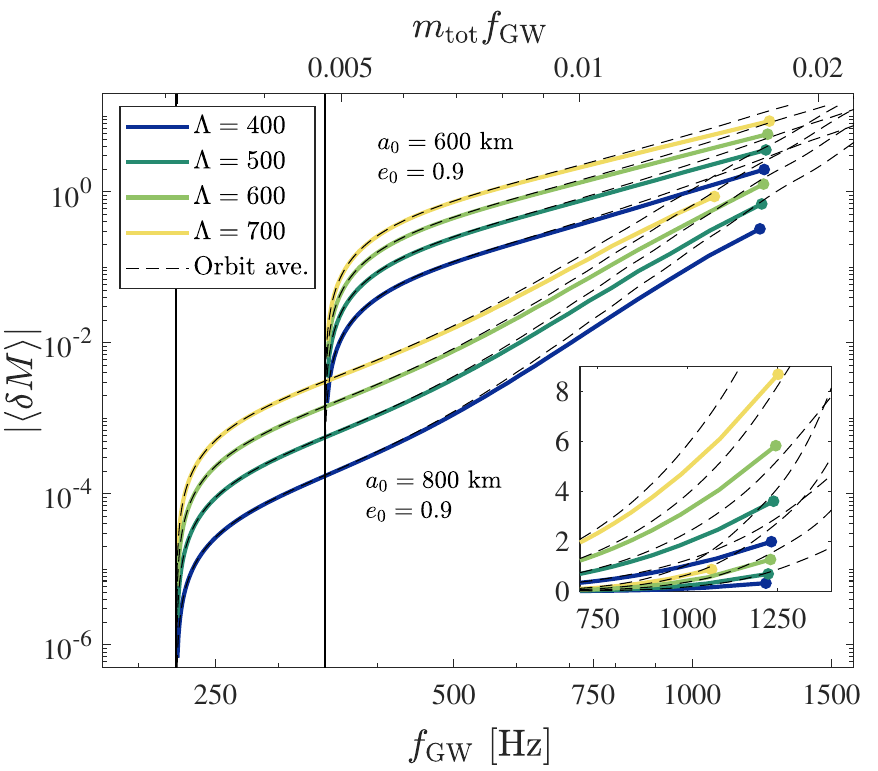}
    \caption{Same as Fig.~\ref{fig:deltaM_an} but with the phase shift $\delta M$ shown as a function of the peak GW frequency $f_{\mathrm{GW}}$ (Eq.~\ref{eq:fGW}). The inset shows the high-frequency part of the evolution on a linear scale.}
    \label{fig:deltaM_fGW}
\end{figure}

An eccentric binary emits GWs in a wide range of frequencies. The GW spectrum of such a binary exhibits a broad peak centered at approximately \cite{Wen:2002km}:
\begin{equation}
    f_\mathrm{GW} = n_\mathrm{peak} f_\mathrm{orb} \approx \frac{2(1+e)^{1.1954}}{(1-e^2)^{3/2}} f_\mathrm{orb} \: .
\label{eq:fGW}
\end{equation}
Note that this formula gives $2f_\mathrm{orb}$ for $e=0$, as is expected for the circular case, while it tends to infinity as $e\to1$. Fig.~\ref{fig:deltaM_fGW} shows the dependence of the phase shift on the peak GW frequency $f_\mathrm{GW}$. Although the initial orbital frequencies of the binary systems shown are below $10$~Hz, the peak GW frequency is above $200$~Hz for both cases due to the high initial eccentricity $e_0=0.9$. This initial peak GW frequency becomes lower as we increase the initial semimajor axis, and so does the magnitude of the phase shift.

\subsection{Comparison to numerical simulations}
\label{ssec:numsim}

As already described in Sec.~\ref{sec:model}, the Newtonian equations of motion were derived from the Lagrangian in Eq.~\eqref{eq:lagrange}. We complemented these equations by adding the dissipative term proportional to $\dot{\boldsymbol{Q}}$, which corresponds to gravitational radiation generated by the f-mode oscillations. Additionally, we have included the 2.5PN terms accounting for the orbital GW radiation back-reaction. In the previous discussions we neglected the conservative 1PN and 2PN corrections to the orbital motion. We examine the effects of these post-Newtonian terms at the end of this section in both the numerical simulations and the analytical expressions derived below. The equations of motion are given in Appendix~\ref{sec:appendix}. The dynamical variables were transformed into a basis where all the equations of motion are real. We solve these equations using a fourth-order Runge-Kutta differential equation solver with adaptive timesteps. The simulations are initialized at the apocenter with initial conditions calculated from the Keplerian parameters, while the initial tidal deformations are set to zero. The simulations are terminated if the separation becomes smaller than $28$~km, which implies direct collision with a conservative NS radius of $14$~km.

To compare the numerical simulation results with the analytical calculations a few more additional steps are needed to extract the shift in the mean anomaly for a given orbital frequency. Both $M$ and $f_\mathrm{orb}$ only exist in the Keplerian picture and require special treatment to determine from the numerical simulations, which also include post-Newtonian and tidal interaction terms. We utilize the fact that the orbit is Keplerian to a good approximation at the apocenter. Therefore, we calculate all the Keplerian parameters at each apocenter passage using the energy and angular momentum of the binary system (see Appendix~\ref{sec:appendix}), from which the orbital frequency can be determined as well. The mean anomaly increases by $2\pi$ at each apocenter passage\footnote{We find that due to numerical inaccuracies, more accurate results can be obtained by directly calculating the mean anomaly from the orbital parameters (see Appendix~\ref{sec:appendix}).}. It was stressed below Eq.~\eqref{eq:ddM} that the analytical formula for the phase shift is meaningful when the orbital evolutionary tracks are compared at the same orbital frequency. However, the orbital frequency for the evolution with and without dynamical tides slightly differ for a given apocenter passage. Hence, in order to obtain the phase shift at a given frequency, we interpolate the mean anomaly between adjacent apocenter passages. 

In addition to dynamical tides, the phase of the binary orbit is also influenced by adiabatic tides, which can be obtained by taking the adiabatic limit, $f_0 \gg f_\mathrm{orb}$. For circular binaries the orbital phase shift introduced by adiabatic tides was calculated by Flanagan \& Hinderer to be $\mathcal{O}(1)$ at $f_\mathrm{orb}\gtrsim100$~Hz \cite{Flanagan:2007ix}, while it has been recently shown that the additional phase shift is $\mathcal{O}(10^{-4}-10^{-2})$ in the adiabatic limit for moderate eccentricities \cite{Bernaldez:2023xoh}. Since the analytic formulas, Eqs.~\eqref{eq:deltaM_an} and \eqref{eq:deltaM_orbave} only account for the extra phase shift caused by dynamical tides, we extract this additional phase shift from our numerical simulations as well. We calculate the excess phase shift caused by dynamical tides by subtracting the phase shift from adiabatic tides
\begin{equation}
    \delta M(f_\mathrm{orb}) = M_{f_0}(f_\mathrm{orb}) - M_{f_\mathrm{ad}}(f_\mathrm{orb}) \: ,
\label{eq:dM_excess}
\end{equation}
where $M_{f_0}$ and $M_{f_\mathrm{ad}}$ denote the mean anomaly for f-mode frequency $f_0$ and  $f_\mathrm{ad}\gg f_\mathrm{orb}$, respectively. For practical purposes it is sufficient to set $f_\mathrm{ad}$ to be a few times $f_0$, due to the exponential suppression of tidal excitations with increasing $f_0$ (see Fig.~\ref{fig:Hansen} and the discussion therein). We obtain $\delta M$ as follows. For each set of initial parameters we run two separate simulations, one with $f_0$ being appropriate for dynamical tides to be relevant, and another initialised by setting $f_0 = f_\mathrm{ad}\gg f_\mathrm{orb}$. Then we subtract the mean anomalies evaluated at the same frequencies. The expectation value of the phase shift evaluated in this way should essentially be the same as the phase shift shown by the solid line in Fig.~\ref{fig:deltaM_an}, since there the orbital parameters $a$ and $e$ were already calculated from the numerical simulation taking into account adiabatic tides as well. The only region where we expect the numerical results to deviate from the analytic calculations is where the energy transmitted to dynamical tides becomes comparable to the energy radiated through GWs, and thus where our perturbative assumption breaks down.

\begin{figure}[t]
    \centering
    \includegraphics[width=0.49\textwidth]{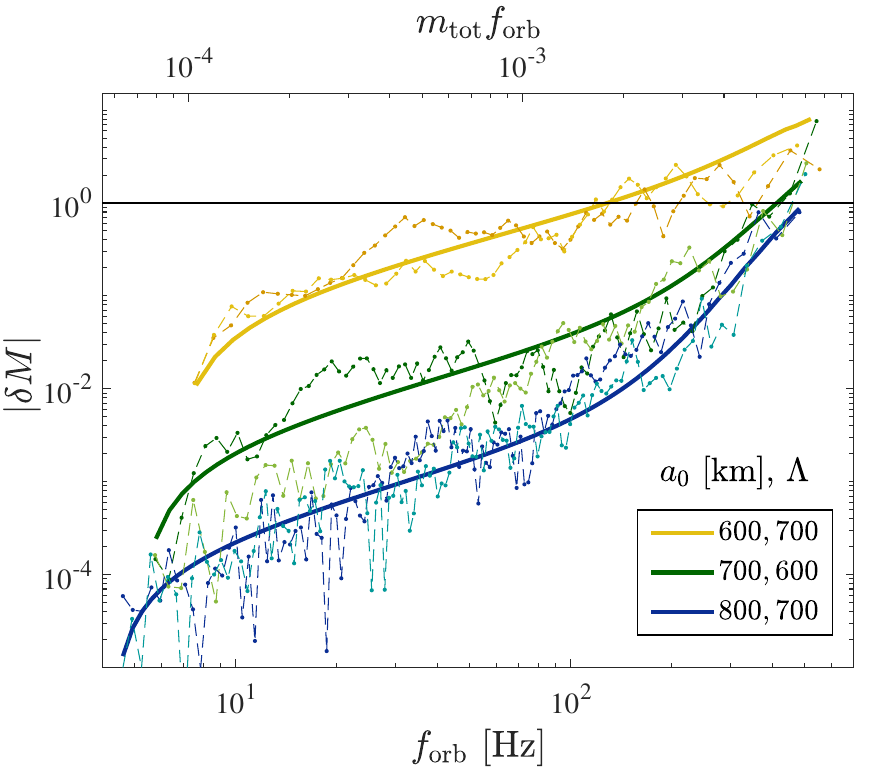}
    \caption{The excess phase shift caused by dynamical tides $\delta M$ as a function of the orbital frequency for binary NSs with two $1.4~\Msun$ components obtained from the numerical simulation (dashed lines) and calculated using the  Eq.~\eqref{eq:deltaM_an} (solid lines). For a given color the two numerical results with different tones correspond to evolutions with slightly different parameter configurations. $e_0=0.9$ for each case.}
    \label{fig:deltaM_num}
\end{figure}

Fig.~\ref{fig:deltaM_num} shows the results obtained from this procedure for binary NSs with two $1.4~\Msun$ components and three different configurations of initial semi-major axis and tidal deformability. The initial eccentricity is $e_0=0.9$ for all cases. The solid lines represent the semi-analytical results obtained from Eq.~\eqref{eq:deltaM_an}, utilizing the numerically calculated orbital evolution without dynamical tides, while the dashed lines represent the numerical results obtained from the procedure described in the previous paragraphs. For each combination of binary parameters, two results from numerical simulations with slightly different parameters are shown. Due to the stochastic motion of phase shifts between consecutive orbits, the difference between the two evolutions can become considerable, however, they both follow the trend given by the semi-analytic formula. Even though the magnitude of the phase shifts for the three different configurations differs by $2-3$ orders of magnitude at low orbital frequencies, their difference shrinks to a single order of magnitude by the merger. The phase shifts approach or even exceed 1 radian by the time the binaries merge for all three configurations.

\begin{figure}[t]
    \centering
    \includegraphics[width=0.49\textwidth]{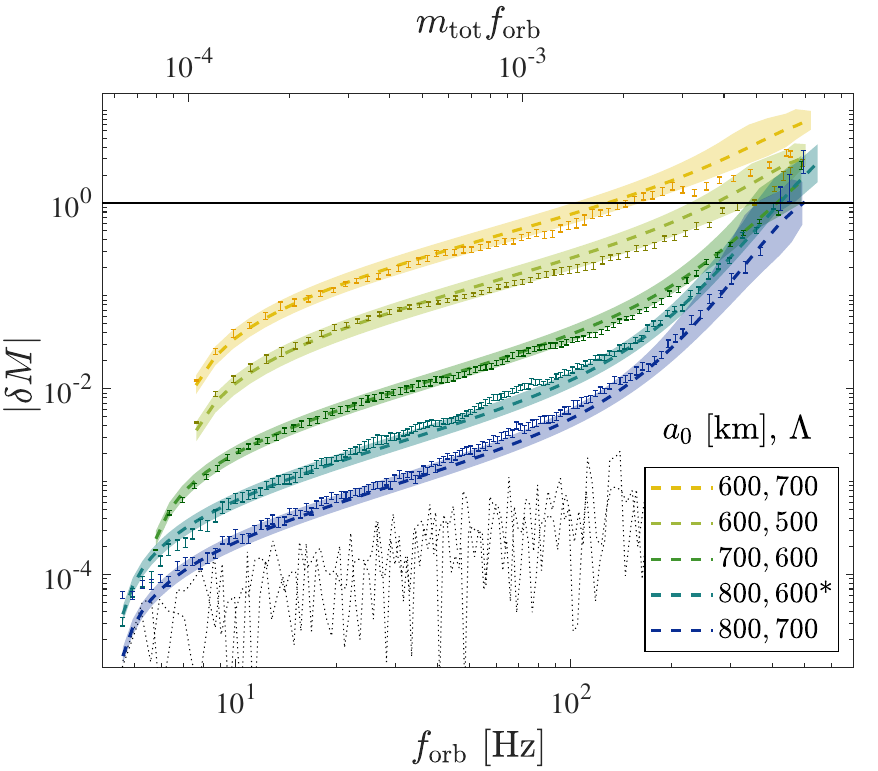}
    \caption{The phase shift $\delta M$ as a function of the orbital frequency for binary NSs with two $1.4~\Msun$ components, averaged for $\sim50$ orbital evolutions with slightly different parameters (error bars). The colored regions show the areas covered by the results calculated from Eq.~\eqref{eq:deltaM_an}, with the dashed lines representing their average. For the results with $a_0=800$~km and $\Lambda=600$, the f-mode frequency was taken to be off the quasi-universal relation, with $f_0=1.5$~kHz. $e_0=0.9$ for each case.}
    \label{fig:deltaM_num_ave}
\end{figure}

Note that a related measurable quantity is the phase shift measured at a specific time, e.g. at each apocenter passage of the orbital evolution without dynamical tides.\footnote{In this case the phase shift would be the difference between the phases of the orbital evolutions with and without dynamical tides, both measured at the same time defined by the apocenter passage in case of no dynamical tides.} Note that this quantity differs from the one shown in Figs.~\ref{fig:deltaM_an}-\ref{fig:deltaM_num_ave}, since orbits with and without dynamical tides reach a given frequency at different times.

Given the stochastic nature of the evolution, particularly the sensitivity of the impulsive tidal kicks at close approaches to the actual phase of the stellar oscillations, we examine the expectation value and scatter of the phase shift for an ensemble of simulations with slightly different initial conditions. This is performed with care given the subtle dependence on the variations of the initial conditions. Interestingly, for non-zero initial deformations, varying the initial phase of the quadrupolar oscillation modes does not change the time intervals between pericenter passages, which ultimately determine the stochastic evolution of the tidal amplitudes for a given $f_0$. The evolution of the mode amplitudes may change by modifying either the f-mode frequency or the orbital evolution through $a_0$ or $e_0$. Also note that variations in $e_0$ and $a_0$ are not independent, changing $e_0$ to $e_0'$ is equivalent to (i.e. results in the same evolutionary path) modifying $a_0$ appropriately to some $a_0'$ while keeping $e_0$ fixed, i.e. along the evolutionary paths shown in Figure~\ref{fig:dE}. Hence, we are left with two independent parameters, e.g. $f_0$ and $a_0$, that can be independently changed in order to get a different evolution for the tidal amplitudes. The variation in these parameters must also be kept sufficiently low in order to simulate approximately the same orbital evolution, but variations chosen to be too small may lead to negligible deviations in the evolution. These constraints limit the range of orbital configurations which can be averaged over.

Fig.~\ref{fig:deltaM_num_ave} shows the averaged values of phase shifts with error bars for different parameter configurations, obtained by averaging the phase shifts from $\sim50$ numerical simulations with slightly different parameter configurations. The regions covered by the phase shifts calculated from the semi-analytical formula Eq.~\eqref{eq:deltaM_an} are indicated by the colored regions, with their average also indicated by the dashed lines. For most parts of the results there is a good agreement between the numerical averages and the analytical results, however, there are a few exceptions. The semi-analytical results overestimate the numerical average close to the merger for the simulations with $a_0=600$~km. This can be attributed to the fact that Eq.~\eqref{eq:deltaM_an} was obtained from a perturbative approach, while the phase shifts at these stages can no longer be considered small perturbations. Another discrepancy can be noticed at the low frequency end of the parameter configuration with $a_0=800$~km and $\Lambda=700$, where the numerical phase shifts are larger than the analytical ones. This is due to numerical error which causes the phase difference to be higher than $10^{-5}-10^{-3}$ depending on the orbital frequency even when they should be lower according to the analytic formulas. To assess the extent of these limitations we calculated the phase difference for a single configuration with the same f-mode frequency, but with two different choices for numerical precision. Since we compare the same evolution to itself, the phase difference should be zero, however the numerical results differ. The results from two different such simulations are shown by the dotted black lines in Fig.~\ref{fig:deltaM_num_ave}. Even though this numerical artefact is well below the observable region and while it is slowly suppressed by increasing the numerical precision, it is not fully understood why it emerges. They might be linked to the frequency interpolation of phases.

The semi-analytical phase shift estimates provide an accurate interpretation of the numerical simulations also for tidal parameters off the quasi-universal relation.

\begin{figure}[t]
    \centering
    \includegraphics[width=0.49\textwidth]{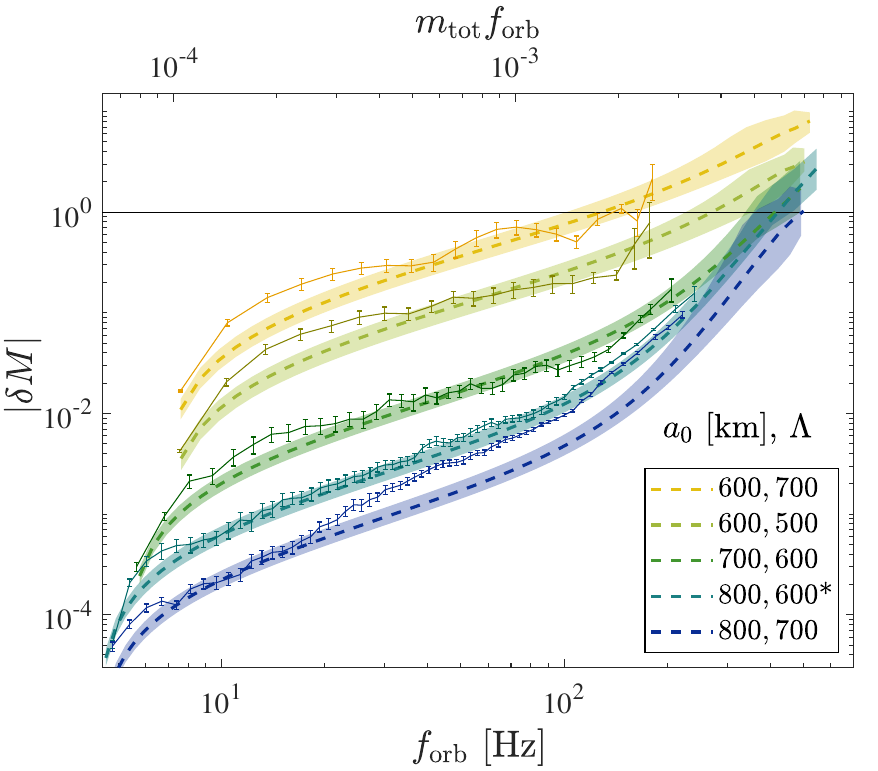}
    \caption{The phase shift $\delta M$ as a function of the orbital frequency for binary NSs with two $1.4~\Msun$ components, averaged for $\sim50$ orbital evolutions with slightly different parameters (error bars). The orbital evolutions were calculated using the 1PN and 2PN terms as well. The colored regions show the areas covered by the results analytically calculated from Eq.~\eqref{eq:deltaM_an} without the 1PN and 2PN terms, with the dashed lines representing their average. For the results with $a_0=800$~km and $\Lambda=600$, the f-mode frequency was taken to be off the quasi-universal relation, with $f_0=1.5$~kHz. $e_0=0.9$ for each case.}
    \label{fig:deltaM_num_PN}
\end{figure}

As we mentioned in the beginning of this subsection, a correct treatment of the post-Newtonian interactions would require taking into account additional corrections to equations of motion used in the numerical simulations and the analytical formulas. For a simple test, we estimate the change in the phase shift evolution caused solely by the 1PN and 2PN terms added to the orbital evolution equations (see Eqs.~\ref{eq:rPN} and \ref{eq:phiPN}. Fig.~\ref{fig:deltaM_num_PN} shows the results for the phase shift as a function of the orbital frequency when the 1PN and 2PN terms are included in the orbital equations of motion. The numerical evaluation of the phase shift is described in Appendix~\ref{sec:appendix}. $f_\mathrm{orb}$ was calculated using the PN corrections described in Ref.~\cite{Blanchet:2013haa}. There is a good overall agreement between the analytic and numerical results, one major difference being that the phase shifts for $a_0=800\,$km and $\Lambda=700$ are systematically increased compared to the analytic results. Since the relative scatter of phase shifts is also reduced in this case, we think that it might be caused by an additional, non-stochastic effect, although the exact reason remains unclear. It should also be noted that with the inclusion of the 1PN and 2PN corrections, the orbital frequency at merger is reduced by about a factor of two, resulting in a lower phase shift at merger, while the total number of orbits is also reduced.

\subsection{Comparison to the literature}
\label{ssec:literature}

Our results differ from the results of Vick and Lai \cite{Vick:2019cun} who studied the GW phase shift in the time domain for binary NS systems. Their Fig.~5 showed spikes in the GW phase shift around each pericenter passage, which exceeded $\sim 10$~radians close to the merger. They also investigated the differences in GW phases measured at the merger of the binary as a function of the initial eccentricity in their Fig.~7. The latter figure showed that the GW phase shift at merger has an oscillatory behaviour as a function of $e_0$, with this phase shift also changing sign at high eccentricities. We believe that these phase shift spikes and the oscillatory $e_0$ dependence are not directly measurable. Fig.~5 in Ref. \cite{Vick:2019cun} was obtained by matching the GW phases at the start of their simulations, but in a realistic scenario the GW signal of the template is matched to minimize the SNR of the difference relative to the measured data. The corresponding relevant quantity to predict the expected significance of the tidal perturbation in a measurement is the difference SNR of the waveforms with and without tidal effects. For practical purposes this approximately corresponds to setting the time-shift to match the waveforms at the maximum strengths (i.e. at pericenter passages close to the merger). This is expected to diminish the spikes seen in Fig.~5 of that paper and the overall phase shift is expected to be reduced. Larger phase differences then appear at lower frequencies, where the GW signal is weaker and the detectors might be much less sensitive. As for Fig.~7 in Ref. \cite{Vick:2019cun}, since the GW phase at merger is highly dependent on the assumptions of the merger, the oscillatory behaviour shown in that figure is very sensitive to the uncertain assumptions of the simulation and may not be a robust indicator of the properties of the real physical system. In fact, the condition $r = 2.5R$ used in the simulations as the merger condition in that paper, with $R$ being the NS radius, is only a rough estimate, similar to the one we used in our simulations. Indeed, since dynamical tides remove energy from the system, the frequency of the orbital evolution with dynamical tides at a given time can only be larger than the frequency of the evolution without dynamical tides. This means that the phase difference will always be positive\footnote{The definition of the phase shift in that paper contains a minus sign, unlike in our definition.}, as is the case in Fig.~5 of Ref.~\cite{Vick:2019cun}. Thus, the negative values shown in Fig.~7 of Ref.~\cite{Vick:2019cun} may be a consequence of comparing the phases of the GW signals at different times based on the arbitrary choice of the merger condition. A more adequate assessment of the phase shift is shown in Fig.~6 of Ref.~\cite{Vick:2019cun}, which shows the cumulative time advance of the orbit with dynamical tides at each pericenter passage relative to the time of pericenter passages without such tides, as a function of the number of consecutive pericenter passages. Here the time advances monotonically increase with each subsequent orbit and the corresponding phase advances of the mean anomaly at a given time can be estimated to be $\mathcal{O}(1)$ near the merger, which is consistent with our results.

It is also worth noting that several other studies consider a different parameterization of the Lagrangian, utilizing the tidal overlap integral $Q_\xi$ (see e.g. Refs.~\cite{Lai:1993di,Ho:1998hq,Vick:2019cun,Yang:2019kmf}). The connection between $\lambda$ and $Q_\xi$ is given by $\lambda\approx 4\pi Q_\xi^2/(15\omega_0^2)$ using the definition of $Q_\xi$ as given in Ref.~\cite{Yang:2019kmf}.

\begin{figure}[t]
    \centering
    \includegraphics[width=0.49\textwidth]{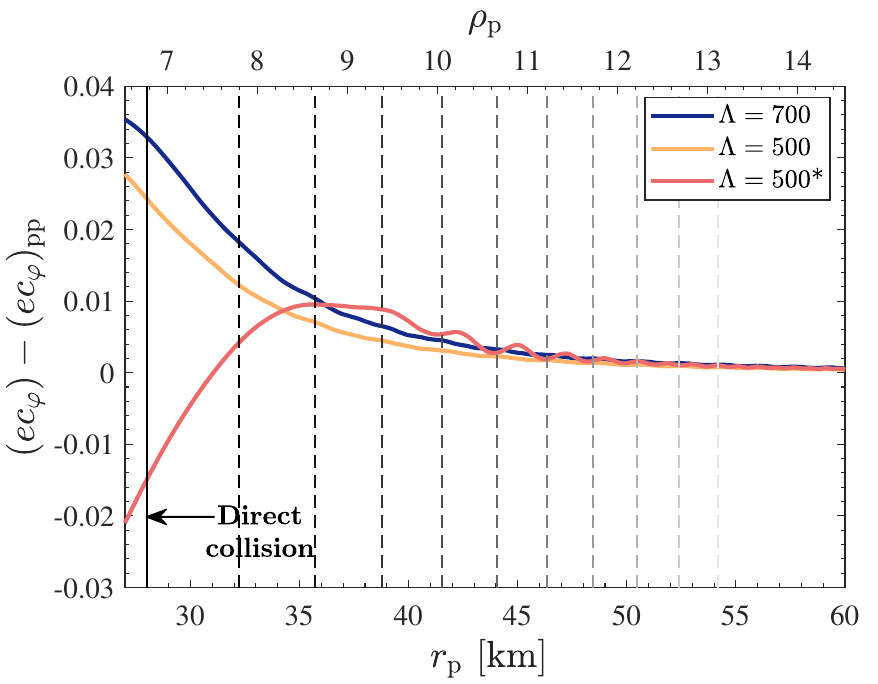}
    \caption{The change in the oscillatory eccentricity $ec_\varphi$ due to tidal interactions compared to the point-particle case. The solid blue and yellow lines correspond to non-spinning NSs with different tidal deformabilities, while the red line correspond to NSs with the f-mode frequency modified to $900$~Hz, representing highly-spinning NSs (with $f_\mathrm{rot}\approx800$~Hz), with anti-aligned spins. Each vertical dashed line represents a single orbit.}
    \label{fig:ecc_from_tide}
\end{figure}

Another interesting phenomenon was shown in Ref.~\cite{Yu:2024uxt}, who found that initially circular binaries can become slightly eccentric due to strong tidal torques during the late inspiral. They showed that for binaries of highly spinning NSs with anti-aligned spins the eccentricity can become as large as $0.05-0.1$ right before the merger. Fig.~\ref{fig:ecc_from_tide} shows results from our simulations on the eccentricity evolution of the binary without the 1PN and 2PN terms. Similarly to Ref.~\cite{Yu:2024uxt} we show the evolution of the osculating orbital element $ec_\varphi$ compared to the point-particle (pp) case, where
\begin{equation}
    ec_\varphi \equiv e \cos(\varphi-\varphi_0) = \frac{p}{r}-1 = \frac{r^3\dot{\varphi}^2}{m_\mathrm{tot}}-1 \:,
\end{equation}
with $\varphi_0$ being the argument of pericenter, and $p$ the semilatus rectum. Our findings are consistent with those of Ref.~\cite{Yu:2024uxt} in that a finite eccentricity is induced by tides right before the merger. We find that this effect is substantial even for non-spinning NSs. However, we note that eccentricities of $e\gtrsim0.01$ are only induced $2-4$ orbits before direct collision of the NSs, where the Newtonian approximation might break down.

\subsection{Detectability of the tidal oscillations}
\label{ssec:detect}
While the orbital phase shift can provide a general guide about the observability of dynamical tides, a more rigorous test of detectability may be obtained by calculating the SNR betwen the difference of the evolutionary curves with and without the dynamical tides. We calculate the $h_{+}$ and $h_{\times}$ components of the GW signal generated by the orbital motion using the quadrupole formula, $h_{ij}=2\ddot{I}_{ij}/d$, where $d$ is the distance of the binary and $I_{ij}$ is the trace-free quadrupole moment. We also calculate the GW signal emitted by the stellar oscillations, $h^Q_+$ and $h^Q_\times $. For GWs propagating perpendicularly to the orbital plane, along the $z$ axis, the two polarizations in the transverse--traceless gauge become:
\begin{equation}
    h_+ = \frac{1}{d}\left( \ddot{I}_{xx} - \ddot{I}_{yy} \right) \: , \quad h_\times = \frac{2}{d} \ddot{I}_{xy} \: ,
    \label{eq:hpol}
\end{equation}
which gives:
\begin{align}
    h_+ = \frac{2\mu}{d} \left( \dot{r}^2 \cos2\varphi + r\ddot{r} \cos2\varphi - 4 r \dot{r} \dot{\varphi} \sin2\varphi \right. \nonumber \\
    - \left. 2r^2\dot{\varphi}^2 \cos2\varphi - r^2\ddot{\varphi} \sin2\varphi \right) \: , \\
    h_\times = \frac{2\mu}{d} \left( \dot{r}^2 \sin2\varphi + r\ddot{r} \sin2\varphi + 4 r \dot{r} \dot{\varphi} \cos2\varphi \right. \nonumber \\
    - \left. 2r^2\dot{\varphi}^2 \sin2\varphi + r^2\ddot{\varphi} \cos2\varphi \right) \: .
\end{align}
The same components for the GW signal induced by the tidal oscillations are
\begin{equation}
    h^Q_+ = \frac{\sqrt{2}}{d} \ddot{Q}_+ \: , \qquad h^Q_\times = \frac{\sqrt{2}}{d} \ddot{Q}_\times \: ,
\end{equation}
with the variables $Q_+$ and $Q_\times$ introduced in Appendix~\ref{sec:appendix}. 

Generally, the SNR can be calculated for any waveform using the Fourier-spectrum of the measured strain, $\tilde{h}(f)$, assuming Wiener-filtering (see e.g. Ref.~\cite{Thrane:2014yza}):
\begin{equation}
    \mathrm{SNR}^2 = 4 \int_0^{\infty} \frac{|\tilde{h}(f)|^2}{S_h(f)} \mathrm{d}f \: ,
\end{equation}
where $S_h(f)$ is the one-sided spectral density of the detector noise\footnote{We used the updated Advanced LIGO sensitivity curve:\\ \href{https://dcc.ligo.org/T1800044-v5/public}{https://dcc.ligo.org/T1800044-v5/public}} and the complicated angular depdence is incorporated in $\tilde{h}(f)$ For L-shaped detectors, like LIGO, Virgo and KAGRA, the position and orientation averaged SNR is reduced by a factor of $2/5$ compared to the SNR in the optimal (face-on) orientation and sky position (see e.g. Ref.~\cite{Dalal:2006qt}):
\begin{equation}
    \mathrm{SNR}_\mathrm{ave}^\mathrm{LIGO} = \frac{2}{5} \mathrm{SNR}_\mathrm{opt}^\mathrm{LIGO} \: ,
\end{equation}
while for the Einstein Telescope this factor is reduced by $\sqrt{3}/2$ due to the $60^\circ$ opening angle between the interferometer arms, but increased by a factor of $\sqrt{3}$ due to essentially acting as 3 independent detectors. Hence the above relation becomes \cite{Regimbau:2012ir}:
\begin{equation}
    \mathrm{SNR}_\mathrm{ave}^\mathrm{ET} = \frac{3}{5} \mathrm{SNR}_\mathrm{opt}^\mathrm{ET} \: .
\end{equation}
Then the detection range of a binary system can be defined as the distance where the averaged SNR is 8:
\begin{equation}
    d_{\rm L, det} \, [\mathrm{Mpc}] = \frac{1}{8}\mathrm{SNR}_\mathrm{ave}(d=1~\mathrm{Mpc}) \: .
\end{equation}

Calculating the detection range for directly measuring NS oscillations, we obtain sub-megaparsec distances for a single aLIGO detector and a few Mpcs for the Einstein Telescope, using design sensitivities, even for the highly excited case of $a_0=600$~km, $\Lambda=700$.

In order to assess the SNR of the phase shift caused by dynamical tides let us proceed similarly to Ref.~\cite{Kocsis:2011dr} and calculate the SNR for the subtracted spectrum:
\begin{equation}
    \delta \tilde{h}(f) = \tilde{h}_\mathrm{dyn}(f) - \tilde{h}_\mathrm{ad}(f) \: ,
    \label{eq:deltah}
\end{equation}
where $\tilde{h}_\mathrm{dyn}(f)$ is the GW spectrum of the orbital evolution with  GW radiation reaction and dynamical tides, and $\tilde{h}_\mathrm{ad}(f)$ is the same for the orbital evolution with GW radiation reaction and only adiabatic tides. To mimic matched filtering we fix the phase of the two signals at their highest peak (technically at the last pericenter passage before the simulations are terminated at $r=28$~km, which conservatively corresponds to a direct collision), conservatively truncate the signals beyond this peak and calculate their Fourier-spectra.\footnote{Note that the difference of the two signals vanishes at the truncation boundary and we have verified that the truncation does not introduce a significant spurious powerlaw spectral tail.} We do not account for the GW signal beyond this point since there the assumed stellar oscillation and dissipation model may break, the NSs may get tidally disrupted or be subject to other non-perturbative effects. Since these limitations may not necessarily only apply strictly at the last orbit, we examine the cumulative signal to noise ratio as a function of the pericenter distance $r_{\rm p}$. Note that $r_{\rm p}$ decreases monotonically during the time-evolution.

\begin{figure}[t]
    \centering
    \includegraphics[width=0.49\textwidth]{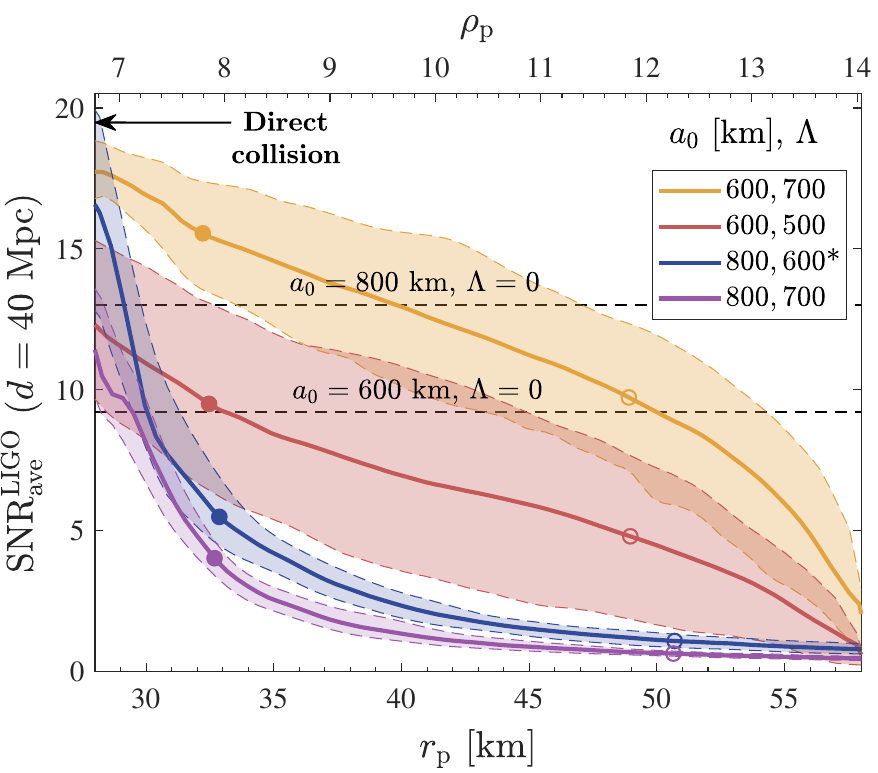}
    \caption{The cumulative signal-to-noise ratio (SNR) of dynamical tides for sources at distance 40 Mpc initialized with $e_0=0.9$ and $a_0$ as labelled until reaching pericenter distance $r_{\rm p}$ shown on the horizontal axis, for different initial conditions and dimensionless tidal deformabilities. Note that direct collision corresponds to $r_\mathrm{p,coll}=28$~km. Solid lines represent the average of individual SNRs, while the coloured patches represent the $68\%$ confidence intervals. Circles correspond to pericenter distances with characteristic GW frequencies (see Eq.~\ref{eq:fGW}) $500$~Hz (empty circle) and $1000$~Hz (filled circle). The horizontal dashed lines show the SNR of the GW signals of orbital evolutions without tidal interactions for the two different initial conditions, respectively. }
    \label{fig:SNR}
\end{figure}

Fig.~\ref{fig:SNR} shows the orientation and sky position averaged cumulative SNR for measuring the tidal effect with a single aLIGO detector as a function of the pericenter distance $r_\mathrm{p}$, calculated from the difference spectrum $\delta\tilde{h}$. The distance of the binaries was chosen to be the same as that of the source of GW170817, which was estimated to be $\sim40$~Mpc \cite{LIGOScientific:2018hze}. The SNR without tidal interactions are also indicated by the horizontal dashed lines for the same initial conditions. As expected from the phase shifts shown in the previous figures, the SNR for detecting dynamical tides can become significant. For the two configurations with $a_0=600$~km ($r_{\rm p0}=60$~km) it exceeds the SNR of the unperturbed orbit. At this point the tidal effects obviously can no longer cosidered to be small perturbations to the original GW-driven orbit plus adiabatic tides, as it is assumed otherwise. The figure shows that the SNR accumulates to detectable levels already at relatively large periapsis separations, where the post-Newtonian treatment and the treatment of the tidal effects may be expected to be valid. The phase shifts for the orbits with $a_0=800$~km ($r_{\rm p0}=80$~km), which experience weaker tidal interactions, also become significant at small separations. Also note that these results only correspond to a single aLIGO detector and considering the whole GW detector network would increase these by a factor of $1.5-2$. This means that it might be feasible to observe the effect of dynamical tides already with the current network of GW detectors. The SNRs (or equivalently the detection ranges) of dynamical tides increase by more than an order of magnitude when we calculate with the design sensitivity of the Einstein Telescope. The detection ranges for the two detectors and the different binary configurations at two different characteristic GW frequencies are listed in Table~\ref{tab:SNR}.

\renewcommand{\baselinestretch}{1.5}
\begin{table}[!tb]
\centering
\resizebox{0.82\width}{!}{$
\begin{tabular}[c]{|c|c|c|c|c|}\cline{2-5}
\multicolumn{1}{c|}{} & \multicolumn{4}{c|}{$a_0$ [km], $\Lambda$} \\\hline
Detection range [Mpc] & 800, 700 &  800, 600* &  600, 500 &  600, 700 \\\hline\hline
$d_{\rm L, det}^\mathrm{LIGO}$ ($\Lambda=0$) & 65 & 65 & 46 & 46 \\\hline
$d_{\rm L, det}^\mathrm{ET}$ ($\Lambda=0$) & 817 & 817 & 629 & 629 \\\hline
$d_{\rm L, det}^\mathrm{LIGO}$ ($f_\mathrm{GW}=500$ Hz) & $3.1^{+0.4}_{-0.5}$ & $4.8^{+1.0}_{-0.9}$ & $25^{+13}_{-15}$ & $43^{+12}_{-14}$ \\\hline
$d_{\rm L, det}^\mathrm{ET}$ ($f_\mathrm{GW}=500$ Hz) & $50.9^{+8.5}_{-8.0}$ & $82^{+13}_{-17}$ & $390^{+260}_{-280}$ & $760^{+220}_{-220}$ \\\hline
$d_{\rm L, det}^\mathrm{LIGO}$ ($f_\mathrm{GW}=1$ kHz) & $17.2^{+1.9}_{-3.3}$ & $26.9^{+5.6}_{-5.3}$ & $46^{+14}_{-17}$ & $77^{+10}_{-12}$ \\\hline
$d_{\rm L, det}^\mathrm{ET}$ ($f_\mathrm{GW}=1$ kHz) & $280^{+34}_{-49}$ & $440^{+77}_{-86}$ & $770^{+280}_{-280}$ & $1270^{+180}_{-180}$ \\\hline
\end{tabular}
$}
\renewcommand{\baselinestretch}{1}
\caption{\label{tab:SNR}Detection ranges for different binary configurations at two different GW frequencies, calculated by using the design sensitivity curves of a single aLIGO detector as well as the Einstein Telescope. The detection ranges for $\Lambda=0$ were calculated from the orbit-averaged results \cite{OLeary:2008myb}.}
\end{table}
\renewcommand{\baselinestretch}{1}

It is also important to point out that due to the strong dependence of $\delta M$ on the f-mode frequency, the phase shift and therefore the SNR of a binary system with $\Lambda = 600$ greatly increases if we set the f-mode frequency to $1.5$~kHz instead of $\sim1.66$~kHz predicted by the quasi-universal relation between $\Lambda$ and $f_0$. They even increase above the values obtained for the binary with $\Lambda=700$ and $f_0\approx1.61$~kHz. Due to this strong dependence on the f-mode frequency the measurement of $\delta M$ could be used to test the quasi-universal relation between $\Lambda$ and $f_0$.

\begin{figure*}
    \centering
    \includegraphics[width=0.45\textwidth]{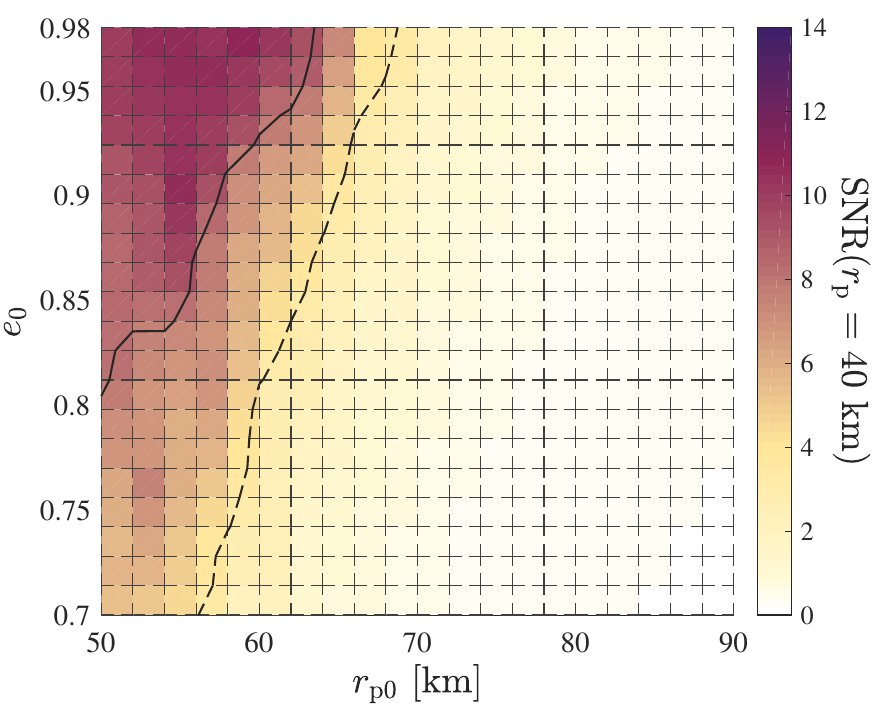}
    \hspace{0.03\textwidth}
    \includegraphics[width=0.45\textwidth]{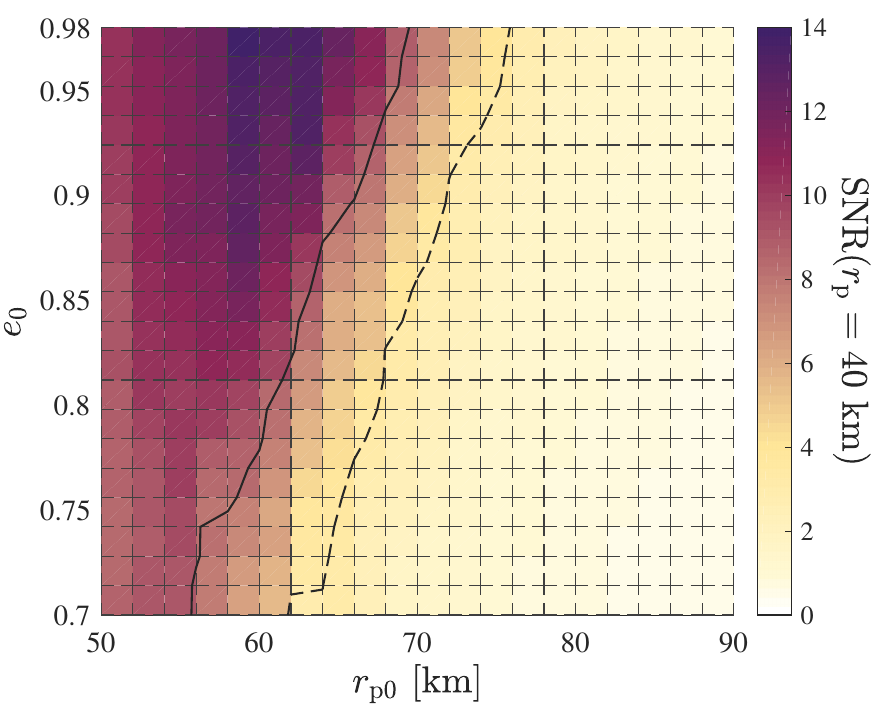}
    \caption{Orientation and sky location averaged signal-to-noise ratio of the subtracted spectrum $\delta h$ (see Eq.~\ref{eq:deltah}), calculated from GW strains truncated at a final periapsis separation of $r_\mathrm{p}=40$~km, as a function of initial orbital parameters $r_\mathrm{p0}$ and $e_0$. The two panels correspond to $\Lambda=500$ (left) and $\Lambda=700$ (right) with corresponding f-mode frequencies of $1.73$~kHz and $1.61$~kHz, respectively. Both panels show values averaged over several, slightly different initial conditions in order to remove stochasticity. The source distance is set to $40$~Mpc, corresponding to the estimated distance of GW170817. The two contour lines correspond to SNRs of $8$ (solid line) and $2/5\cdot 8$ (dashed line), indicating the detectability threshold of a single aLIGO detector with random orientation, and one with optimal orientation, respectively.}
    \label{fig:SNRgrid}
\end{figure*}

Figure~\ref{fig:SNRgrid} shows the orientation and sky location averaged SNR of the subtracted spectrum $\delta h$ (Eq.~\ref{eq:deltah}), calculated from GW strains truncated at $r_\mathrm{p}=40$~km, as a function of initial orbital parameters $r_\mathrm{p0}$ and $e_0$. The distance of the binary is set to $40$~Mpc. Fig.~\ref{fig:SNRgrid} shows that the SNR is greatly increased in the highly eccentric case, compared to the circular or moderately eccentric case (the SNR is increased by a factor of $\sim2$ between $e_0=0.7$ and $e_0=0.98$). While increasing the initial pericenter distance quickly suppresses the effect of dynamical tides, the SNR can still reach the detectability threshold (SNR$\sim8$) for binaries with $r_\mathrm{p0}\lesssim68$~km ($r_\mathrm{p0}\lesssim76$~km) and a tidal deformability parameter of $\Lambda=500$ ($700$), for a single optimally oriented aLIGO detector.

However we note that the assumed detectability limit of SNR$=8$ is usually assumed for GW sources with available matched filtering templates, but it is unclear whether a matched filtering search is feasible for NS binaries with dynamical tides. The detection and measurement of NSs with dynamical tides may be more difficult.

\subsection{Implications for the equation of state}

While dynamical tides affect the evolution of circular binaries as well, they only become significant very close to the merger \cite{Hinderer:2016eia,Schmidt:2019wrl}. Therefore, the tidal phase shift is mostly determined by adiabatic tides, which, to the leading 5PN order, only depend on the mass-weighted tidal deformability:
\begin{equation}\label{eq:tildeLambda}
    \tilde{\Lambda} = \frac{16}{13} \frac{(m_1+12m_2)m_1^4\Lambda_1 + (m_2+12m_1)m_2^4\Lambda_2}{(m_1+m_2)^5} \: .
\end{equation}
Hence, the measurement of adiabatic tides only provide information about this specific combination of binary parameters to the leading 5PN order, while $\Lambda_1$ and $\Lambda_2$ could only be constrained by imposing additional constraints on the EoS. On the other hand, dynamical tides also carry information about the f-mode frequency of tidal oscillations. This could resolve the degeneracy between the binary parameters and could be used together with quasi-universal relations to determine $\Lambda_1$ and $\Lambda_2$ independently, thus also providing information about the radii, $R_1$ and $R_2$ of the two NSs.

An interesting open question about the EoS of dense strongly interacting matter is whether it contains a strong first-order phase transition at densities available for NSs \cite{Annala:2019puf,Takatsy:2023xzf}. In case it does, a new stable branch of compact stars could emerge, containing some form of exotic matter (e.g. quark matter) inside their cores. The two branches are then separated by an unstable branch, characterized by decreasing NS mass with increasing central energy density. Thus, two compact stars having the same mass but positioned on different stable branches, therefore having different radii, can exist, referred to as \emph{twin stars} \cite{Glendenning:1998ag,Alford:2015gna,Christian:2021uhd}.

\begin{figure}[!tb]
    \centering
    \includegraphics[width=0.49\textwidth]{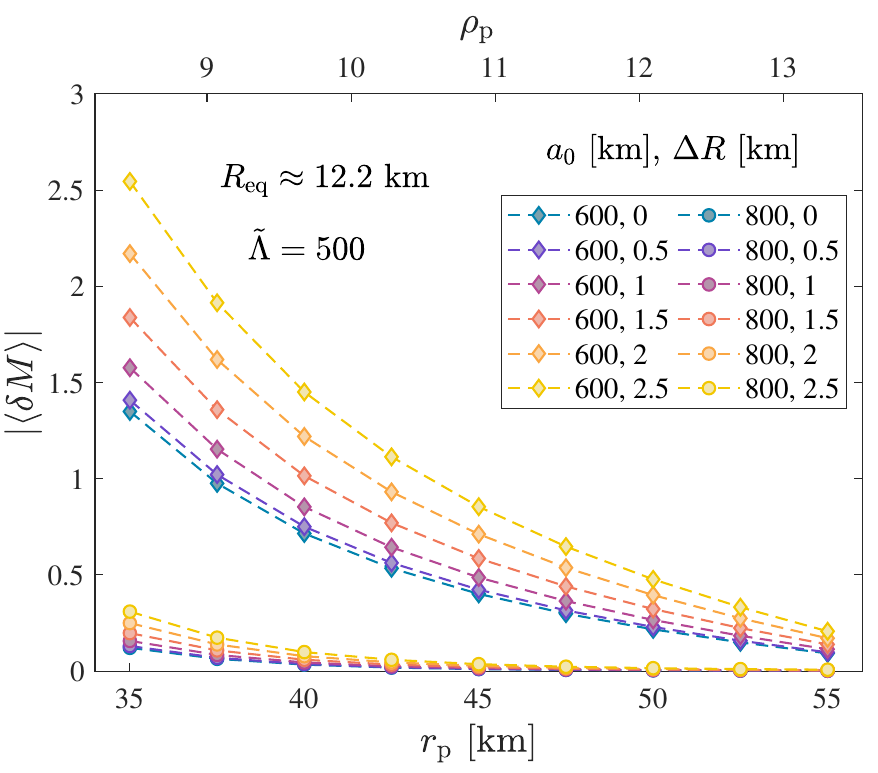}
    \caption{The phase shift at different peak GW frequencies for twin stars (see text) with masses of $1.4~\Msun$ but different radii, resulting in different tidal deformabilities and f-mode frequencies. The phase shifts were obtained from Eq.~\eqref{eq:deltaM_an}, where the input orbital parameters were extracted from our numerical simulations run in the adiabatic tide limit, i.e. with $f_0 = f_\mathrm{ad} \gg f_\mathrm{orb}$. The mass-weighted tidal parameter is kept constant ($\tilde{\Lambda}=500$), while different colors correspond to various differences between the radii of the twin stars. The two different markers correspond to different initial orbital configurations, where $e_0=0.9$ for each case.}
    \label{fig:pt_dR}
\end{figure}
\begin{figure}[!tb]
    \centering
    \includegraphics[width=0.49\textwidth]{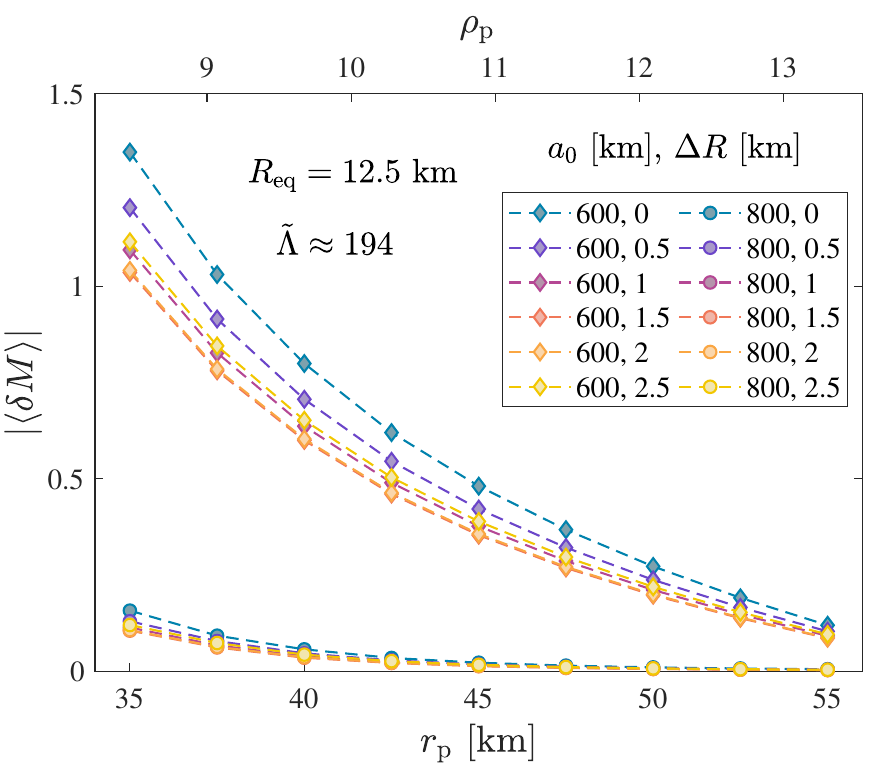}
    \caption{The phase shift at different peak GW frequencies for binary NSs with a $1.4\Msun$ and a $2\Msun$ component. The phase shifts were obtained from Eq.~\eqref{eq:deltaM_an}, where the input orbital parameters were extracted from our numerical simulations run in the adiabatic tide limit, i.e. with $f_0 = f_\mathrm{ad} \gg f_\mathrm{orb}$. The mass-weighted tidal parameter is kept constant ($\tilde{\Lambda}\approx194$, calculated from the equal-radius case with $R_\mathrm{eq}=12.5$~km), while different colors correspond to different slopes for the $M-R$ relation resulting in various differences between the radii of the two NSs ($R(2.0~\Msun)\geq R(1.4~\Msun)$). The two different markers correspond to different initial orbital configurations, where $e_0=0.9$ for each case.}
    \label{fig:Rslope}
\end{figure}

With only the mass-weighted tidal deformability, Eq.~\eqref{eq:tildeLambda}, we cannot distinguish between a binary system consisting of twin stars and one comprised of identical NSs from a single stable branch. On the other hand, measuring the effect of dynamical tides might enable us to break the degeneracy and to differentiate between these two possibilities.

Fig.~\ref{fig:pt_dR} shows the phase shift caused by dynamical tides as a function of pericenter separation for different $a_0$ and binaries with the same component mass $m_1=m_2=1.4\Msun$ but different NS radii with $\Delta R=R_1-R_2$ as shown in the legend. Lines with different colors correspond to binary NSs having the same mass-weighted tidal deformability but different values for the individual tidal deformabilities, and hence different radii. The mass of both NSs was set to $1.4~\Msun$, while we also utilized the quasi-universal relation of Ref.~\cite{Godzieba:2020bbz} between the compactness $C$ and $\Lambda$, in order to determine the individual radii of NSs given a fixed value for $\tilde{\Lambda}$. The f-mode frequencies were determined from the quasi-universal relation of Ref.~\cite{Pradhan:2022vdf} between $\Lambda(1.4~\Msun)$ and $f_0(1.4~\Msun)$. The value of the mass-weighted tidal deformability was set to $\tilde{\Lambda}=500$, which corresponds to a radius of $R_\mathrm{eq}\approx12.2$~km for two NSs with identical radii. Increasing the value of $\Delta R = R_1-R_2$ increases the effect of dynamical tides. For larger values of $\Delta R$ the respective phase shift contribution becomes dominated by the NS with the larger radius, hence the one with the larger tidal deformability and lower f-mode frequency. A radius difference of $2.5$~km increases the phase shift by a factor of $1.5-2$ for $a_0=600$~km and $e_0=0.9$, and by a factor of $2.5-3$ for a binary with $a_0=800$~km and $e_0=0.9$. Smaller radius differences are more difficult to identify using the GW phase shift measurement; a radius difference of $1$~km results in a $15-20\%$ and a $25-45\%$ increase in the phase shift relative to that of a binary with identical stellar radii, respectively. Note that this difference might still be possible to detect, provided a signal with sufficiently large SNR.

Similarly, the slope of the mass--radius curve could also be measured through observations of non-equal mass NS binaries. Similar to Fig.~\ref{fig:pt_dR}, Fig.~\ref{fig:Rslope} shows the phase shift for a binary with a $1.4~\Msun$ and a $2~\Msun$ component. Here, the f-mode frequency and the tidal deformability of the $2~\Msun$ NS was determined from the relations in Refs.~\cite{Pradhan:2023zmg,Godzieba:2020bbz} between the compactness and $f_0$, and the compactness and $\Lambda$, respectively. The radius of the two binary components were chosen so that $R_{2.0\Msun}\geq R_{1.4\Msun}$, consistently with recent EoS constraints \cite{Miller:2021qha,Takatsy:2023xzf,Koehn:2024set}. In the equal-radius case the radii were set to be $R_\mathrm{eq}=12.5$~km, while in all other cases the radii were determined in a way to keep the mass-weighted tidal deformability fixed ($\tilde{\Lambda}\approx194$). An interesting trend can be observed in Fig.~\ref{fig:Rslope}, since the phase shift first decreases with increasing $\Delta R$, then it starts to increase, resulting in the radius differences of $1$~km and $2.5$~km having almost the same effect. This is due to the increase of $f_{0}(1.4~\Msun)$ first dominating and later the decrease of $f_{0}(2~\Msun)$ taking over. The largest difference in the phase shift is $25-30\%$ and $30-40\%$ for the two binary configurations with $a_0=600$~km and $800$~km, respectively.

It needs to be added here that on top of the 5PN leading-order contributions, higher-order terms also contribute to adiabatic tides. The next-to-leading-order contribution to the GW phase shift comes from the 6PN effects, which scale with $\delta\psi\propto \delta\Lambda \omega_\mathrm{orb}^{7/3}$ \cite{Vines:2011ud,Vines:2010ca,JimenezForteza:2018rwr}, where
\begin{align}\label{eq:dLambda}
    \delta \Lambda &= \left( \frac{5095}{28} - \frac{15895}{28\chi_1} + \frac{5715\chi_1}{14} - \frac{325\chi_1^2}{7} \right) \chi_1^5 \Lambda_1 \nonumber \\
    & + (1\leftrightarrow2) \: ,
\end{align}
and $\chi_A = m_A/m_\mathrm{tot}$. While the phase shift induced by the 6PN term is approximately an order of magnitude lower than that of the 5PN term, it still reaches values of $\mathcal{O}(0.1-1)$ during the late inspiral, making it comparable to the dynamical tides considered in this work close to merger. Furthermore, since the phase shift is proportional to $\delta\Lambda$, which is linearly independent of $\tilde{\Lambda}$ as a function of $m_1$ and $m_2$, one might also hope to break the degeneracy between $\Lambda_1$ and $\Lambda_2$ by measuring the 6PN effect. However, note that for a binary NS with $m_1=m_2=m$, the 5PN and 6PN tidal coefficients (Eqs.~\ref{eq:tildeLambda} and \ref{eq:dLambda}) simply become $\tilde{\Lambda}=\Lambda_1+\Lambda_2$ and $\delta\Lambda\propto(\Lambda_1+\Lambda_2)=\tilde{\Lambda}$, respectively. Consequently, for equal-mass binaries, the adiabatic tidal contributions remain degenerate at least up to the 6PN order for binaries with the same $\tilde{\Lambda}$ value. In the non-equal mass case shown in Fig.~\ref{fig:Rslope} for the same values of tidal deformabilities, we find that the largest relative difference in the 6PN phase shift is $\sim1\%$, corresponding to a phase difference of $\mathcal{O}(10^{-3}-10^{-2})$. Therefore, we conclude that 6PN contributions are typically insufficient to break the degeneracy in the measurement of tidal deformabilities with the current sensitivity of GW detectors.

\section{Conclusion}
\label{sec:summary}

While dynamical tides only become relevant during the last couple of orbits for circular inspirals, orbital eccentricity can increase these effects considerably due to the fact that the periapsis distance may be much smaller for a much larger number of periods for highly eccentric orbits. In the eccentric case the spectrum of the driving tidal forces extend to higher orbital harmonics. In this study, we have utilized a post-Newtonian model with realistic NS parameters to investigate the effect of dynamical tides in eccentric NS binary systems.

By calculating the energy transferred to tidal oscillations relative to the amount of energy radiated through GWs we have found a strong, exponential dependence on the dimensionless pericenter distance, with dynamical tides being strongly suppressed for binaries formed with large initial pericenter distances (Figs.~\ref{fig:dE}--\ref{fig:dEBH}). We have shown that the energy transferred to tidal oscillations by dynamical tides are below $10^{-3}$ relative to the energy loss due to GWs for NS-BH binaries in most cases (Fig.~\ref{fig:dEBH}). For NS-NS binaries this ratio increases beyond $10\%$ for periapsis distance of $r_{\rm p}=28~\mathrm{km}-40~\mathrm{km}$ depending on eccentricity.

While GWs generated by tidal oscillations driven by dynamical tides typically cannot be detected directly, their effect on the orbital evolution may be observed through a phase shift in the GW signal of the inspiral. Since the decay timescale of the tidal oscillations is much larger than the orbital period, the amplitude of the oscillations varies stochastically from one orbit to another, causing stochastic variations in the orbital phase shift as well. We have demonstrated that the expectation value of these phase shifts are accurately modelled analytically for the Newtonian model complemented with the leading-order radiation terms (Fig.~\ref{fig:deltaM_an}). We have also shown that the phase shifts calculated from the analytical formula agree well with numerical calculations using our pseudo-Newtonian model (Fig.~\ref{fig:deltaM_num_ave}). We have also found that including the 1PN and 2PN terms in the equations of motion does not change the frequency-dependent evolution of the phase shift significantly, although the orbital frequency before merger is reduced, and therefore so is the phase shift before merger as well (Fig.~\ref{fig:deltaM_num_PN}).

We have shown that the SNR associated with the GW phase shift caused by the dynamical tides can be significant for NSs with an initial eccentricity of 0.9 with periapsis distances of 60-80km at similar distances to the detected source of GW170817 (40~Mpc) even with a single aLIGO detector (Fig.~\ref{fig:SNR}), and the detection range of the dynamical tides for ET lies between a few hundred Mpc to $\sim1$~Gpc for binary NSs with initial pericenter distances $r_\mathrm{p0}\lesssim 80$~km and tidal deformabilities in the reasonable range $\Lambda\sim400-700$ (see Table~\ref{tab:SNR}). Below this range of tidal deformability, the f-mode frequency is larger which suppresses dynamical tides making these phase shifts more difficult to observe. These eccentric sources form in the LIGO frequency band and have a signal to noise ratio well above 8 for aLIGO and may be expected to be detactable if neglecting the perturbations of dynamical tides \cite{Gondan_Kocsis2019,Gondan:2020svr}. However dynamical tides imprint stochastic variations on the orbital phase, particularly at periapsis passages, which may make it more difficult to detect these sources with matched filtering.

Currently it is not well understood how likely it may be to detect highly eccentric neutron star binaries in the Universe. These systems may form during GW captures in dense, high velocity dispersion environments \cite{OLeary:2008myb,Gondan:2017wzd,Gondan:2020svr} or GW captures during binary-single or binary-binary interactions \cite{Samsing:2013kua,Samsing:2017xmd,Zevin:2018kzq} especially in flattened configurations such as in active galactic nuclei (AGN) accretion disks \cite{Samsing:2020tda}. The initial periapsis distribution of such highly eccentric binaries with $e_0\sim 0.99$ peaks at $r_{p0}=100\mathrm{km} (m/1.4\Msun) (v/500\mathrm{km/s})^{-4/7}$, where for GW captures $v$ is the escape velocity from the system and for binary-single interactions it is the initial orbital velocity of the binary \cite{Gondan:2020svr,Kocsis2022}. In comparison, we found that the signal-to-noise ratio of detecting dynamical tides in the GW signal is significant below an initial periapsis of $r_{p0}\lesssim 70$km (Fig.~\ref{fig:SNRgrid}). Thus the existence of highly eccentric NS binaries with strong dynamical tides hinges on the existence of dense clusters abundant in neutron stars with high escape velocities or systems with very tight binaries undergoing scattering interactions. The rate of the former is theoretically expected to be low \citep{Tsang2013}; the rates in AGN may possibly be more substantial \citep{Tagawa:2020jnc,Samsing:2020tda}.

NS seismology, specifically the measurement of the f-mode frequency of NSs in this case, can be immensely useful in characterising the EoS of dense strongly interacting matter, since it can serve as independent and therefore complementary information to measurements of masses, radii and tidal deformability, all characterising the structure of NSs in different ways. While these properties can be connected through quasi-universal relations, deviations from these relations are possible, especially for equations of state with a strong first-order phase transition. We have shown that deviations from these quasi-universal relations can cause significant differences in the GW phase shift due to the strong dependence of these effects on the f-mode frequency of the NSs, which in turn allows us to measure these deviations.

Another interesting possibility provided by f-mode frequency measurements is the ability of telling equal-radius and twin-star binaries apart (i.e. binaries consisting of NSs with the same mass but different radii). We have shown that while there is a complete degeneracy between equal-mass binary NSs with the same or different component radii with respect to the the mass-weighted tidal deformability, characterizing the magnitude of adiabatic tides, this degeneracy is broken by dynamical tides and the two branches can be distinguished by measuring the corresponding phase shift (Fig.~\ref{fig:pt_dR}). We have also shown that dynamical tides might also allow us to measure the slope of the mass--radius relation with unequal-mass binaries, although this effect was found to be much smaller (Fig.~\ref{fig:Rslope}). 

While we do not expect major deviations from our leading-order calculations with respect to the magnitude of the phase shift caused by dynamical tides, an improvement on our assumptions would be necessary in order to be able to build accurate waveform models. First of all, the effect of higher-order post-Newtonian terms modify the tidal excitation spectrum near the merger, which have been neglected in our analytical model. In addition, while back-reaction forces from GW radiation are expected to be accurate for the orbital evolution, as well as for the freely-oscillating part of the tidal evolution, cross-terms can become important near the pericenter passages, hence these interactions should also be revisited. We note however, that in our calculations the quadrupole moments $Q_m$ always remained smaller than $\sim10\%$ of the unperturbed moment of inertia of the NSs, even for the strongest tides, while they were below $\sim1\%$ a few orbits before the merger. Therefore we consider our results to be reliable from the point of view of parametric instabilities or other non-perturbative effects that can lead to the saturation of tides. While the spin frequency of inspiralling NSs is expected to be low and usually aligned with the orbital angular momentum, due to the formation and the reduced inspiral timescale of eccentric binaries, the existence of binary NSs with considerable anti-aligned spins might be possible. In this case, the spin of the NS can effectively reduce the resonance frequency of the f-mode oscillations, leading to even stronger dynamical tides. This scenario should also be investigated in a follow-up study.

\appendix

\section{The gravitational potential}
\label{sec:app_grpot}

Let us define the external and induced Newtonian multipolar moments by first expanding the external gravitational potential (which is non-singular in the origin) in a Taylor-series:
\begin{align}
 \Phi_{\rm ext}({\bf r}) &= \Phi_{\rm ext}(0) + { r}_{i}\partial_{i}\Phi_{\rm ext}\bigr\rvert_0 \nonumber \\
 &+ \frac{1}{2} {r}_{i}{r}_{j}\partial_{i}\partial_{j}\Phi_{\rm ext}\bigr\rvert_0 + \ldots \: .
  \label{UExt}
\end{align}
Since the second term simply generates an overall translation, it can be transformed out. Then, let us write equivalently:
\begin{align}
 \Phi_{\rm ext}({\bf r}) = \Phi_{\rm ext}(0) 
 + \frac{1}{2} {r}_{i}{r}_{j}\mathcal{E}_{ij} + \ldots \: .
  \label{UExtE}
\end{align}

Similarly, one can expand the potential generated by the massive object under the influence of the external potential (which can be singular in the origin but vanishes at infinity) in a multipole-series:
\begin{equation}
 \Phi_{\rm ind}({\bf r}) = -\left(\frac{m}{r} +
  \frac{3}{2}Q_{ij}\frac{{r}_{i}{r}_{j}}{r^{5}} + \ldots\right) \quad ,
  \label{UInd}
\end{equation}
where the dipole moment has been removed by choosing the center-of-mass frame as the reference frame. Using this convention for the quadrupole moment and Eq.~\eqref{UExtE} the interaction energy becomes
\begin{equation}
    U_\mathrm{int} = \frac{1}{2} Q_{ij} \mathcal{E}_{ij} \: ,
\end{equation}
where we have used the trace-free property of $\mathcal{E}_{ij}$.

\section{Equations of motion}
\label{sec:appendix}

For our numerical simulations we find it convenient to introduce the variables $Q_+$ and $Q_\times$ defined by
\begin{equation}
    Q_+ = \frac{Q_{-2}+Q_2}{\sqrt{2}} \: , \quad Q_\times = \frac{Q_{-2}-Q_2}{i\sqrt{2}} \: .
\end{equation}
This way our equations of motion will become purely real. Then, considering only the tidal deformation of object $1$ (similar equations hold for object $2$), the complete equations of motion including 1PN, 2PN and the 2.5PN leading order radiation terms for the orbital evolution are
\begin{eqnarray}
    {\ddot Q}_+ + 2\gamma_0 {\dot Q}_+ + \omega_0^2 Q_+ &=& {3 \over \sqrt{2}}\lambda_1 \omega_0^2 {m_2 \over r^3} \cos2\varphi \ ,\\
    {\ddot Q}_\times + 2\gamma_0 {\dot Q}_\times + \omega_0^2 Q_\times &=& {3 \over \sqrt{2}}\lambda_1 \omega_0^2 {m_2 \over r^3} \sin2\varphi \ ,\\
    {\ddot Q}_0 + 2\gamma_0 {\dot Q}_0 + \omega_0^2 Q_0 &=& \sqrt{3 \over 2}\lambda_1 \omega_0^2 {m_2 \over r^3} \ ,
\end{eqnarray}
for the tidal variables of object $1$ and
\begin{eqnarray}
    {\ddot r} &=& r {\dot \varphi}^2 - \frac{9}{2\sqrt{2}} \frac{m_2}{\mu r^4} \left( Q_+ \cos2\varphi + Q_\times \sin2\varphi + \frac{Q_0}{\sqrt{3}} \right) \nonumber\\
    &-& \frac{m_\mathrm{tot}}{r^2}\left(1+ A_{\rm PN}+A_{5/2}+B_{\rm PN}\dot{r} + B_{5/2}\dot{r}\right)\ , \label{eq:rPN} \\
    {\ddot \varphi} &=&  -2 \frac{\dot{\varphi} \dot{r}}{r} - {3 \over \sqrt{2}} \frac{m_2}{\mu r^5} \left( Q_+ \sin2\varphi - Q_\times \cos2\varphi \right) \nonumber \\
    &-& \frac{m_\mathrm{tot}}{r^2}\dot{\varphi}\left(B_{\rm PN}+B_{5/2}\right)\ \label{eq:phiPN} ,
\end{eqnarray}
for the orbital parameters, with the coefficients \cite{Lincoln:1990ji}
\begin{align}
		A_{5/2} &= -\frac{8\mu}{5r}\dot{r}\left(18v^2 +\frac{2m_\mathrm{tot}}{3r} - 25\dot{r}^2\right),\\
		B_{5/2} &=\ \frac{8\mu}{5r}\left(6v^2 -\frac{2m_\mathrm{tot}}{r} - 15\dot{r}^2\right),\\
		A_{\rm PN} &=\ (1+3\eta)v^2-2(2+\eta)\frac{m_\mathrm{tot}}{r}-\frac{3}{2}\eta\dot{r}^2 \nonumber \\
            &+ \frac{3}{4}(12+29\eta)\left(\frac{m_\mathrm{tot}}{r}\right)^2 + \eta(3-4\eta)v^4 \nonumber \\
		& + \frac{15}{8}\eta(1-3\eta)\dot{r}^4 - \frac{3}{2}\eta(3-4\eta)v^2\dot{r}^2 \nonumber \\
            &- \frac{1}{2}\eta(13-4\eta)\frac{m_\mathrm{tot}}{r}v^2 \nonumber \\
            &- (2+25\eta+2\eta^2)\frac{m_\mathrm{tot}}{r}\dot{r}^2,\\
		B_{\rm PN} &= -2(2-\eta)\dot{r} - \frac{1}{2}\dot{r}\left[\vphantom{\frac{m_\mathrm{tot}}{r}}\eta(15+4\eta)v^2 \right. \nonumber \\
            &\left. -(4+41\eta+8\eta^2)\frac{m_\mathrm{tot}}{r} - 3\eta(3+2\eta)\dot{r}^2\right],
\end{align}
where $v^2 = \dot{r}^2 + r^2\dot{\varphi}^2$, $m_\mathrm{tot}=m_1+m_2$ and $\mu = m_1m_2/m_\mathrm{tot}$ are the total and reduced masses, and $\eta=\mu/m_\mathrm{tot}$ is the symmetric mass ratio. Note that the 1PN and 2PN terms are neglected (i.e. $A_\mathrm{PN} = B_\mathrm{PN} = 0$) during most of our analysis, and only taken into account at the end of Sec.~\ref{ssec:numsim}.
We solve these equations using a fourth-order Runge-Kutta differential equation solver with adaptive timesteps.
\begin{figure}[t]
    \centering
    \includegraphics[width=0.48\textwidth]{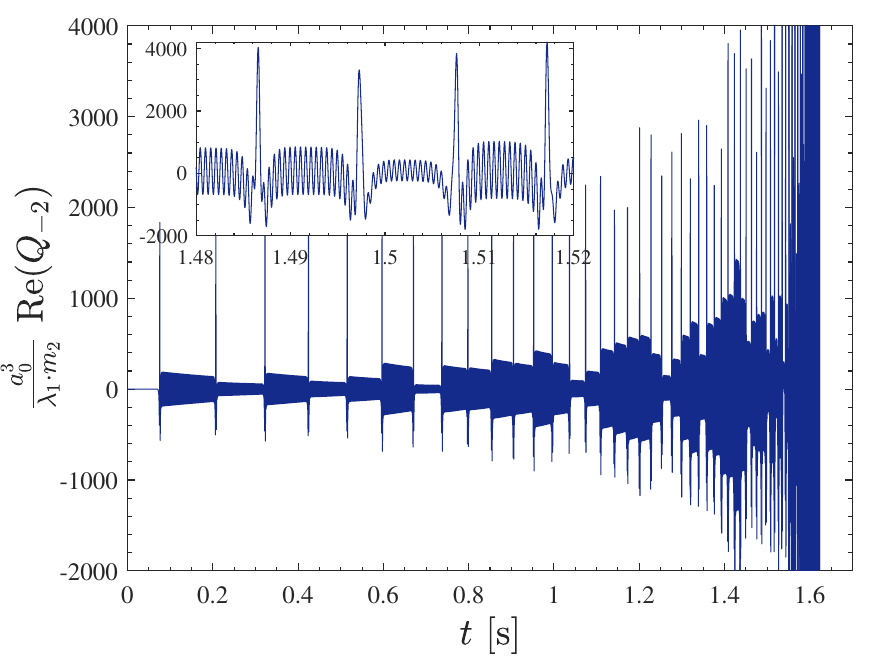}
    \caption{The time evolution of the quadrupole moment Re($Q_{-2}$) for a binary NS with initial orbital parameters $a_0=600$~km and $e_0$, and tidal parameters $\Lambda=400$ and $f_0=1.8$~kHz.}
    \label{fig:Q_2_sim}
\end{figure}

\begin{figure}[t]
    \centering
    \includegraphics[width=0.48\textwidth]{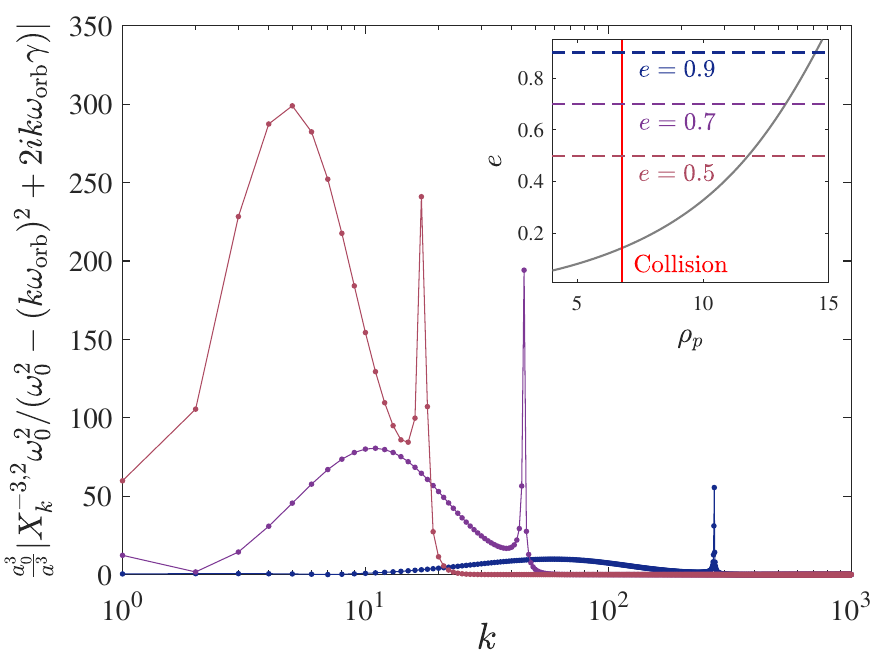}
    \caption{The analytic spectrum of the excited quadrupole oscillation $Q_{-2}$ at different stages of the binary evolution. The initial orbital parameters are $a_0=600$~km and $e_0$, while tidal parameters were chosen as $\Lambda=400$ and $f_0=1.8$~kHz. The datapoints in the discrete spectra are connected to guide the eye. The inset shows the leading-order evolution of the binary parameters without tidal interactions, calculated from Eq.~\ref{eq:aevolution}.}
    \label{fig:Q_spectrum}
\end{figure}

Fig.~\ref{fig:Q_2_sim} shows the typical time evolution of the quadrupole moment. The stochastic motion of the oscillation amplitude from orbit to orbit can be clearly observed. Fig.~\ref{fig:Q_spectrum} shows the analytic spectrum calculated from Eq.~\ref{eq:Qanalytic}, at different stages of the binary evolution.

When $A_\mathrm{PN} = B_\mathrm{PN} = 0$, the Keplerian orbital parameters are calculated at the apocenter from the specific orbital energy and angular momentum:
\begin{align}
    E_\mathrm{orb} = \frac{v_\mathrm{a}^2}{2} - \frac{m_\mathrm{tot}}{r_\mathrm{a}} \: &, \quad L_\mathrm{orb} = r_\mathrm{a}^2 \dot{\varphi}_\mathrm{a} \: , \nonumber \\
    e = \sqrt{1+\frac{2E_\mathrm{orb} L_\mathrm{orb}^2}{m_\mathrm{tot}^2}} \: &, \quad a = -\frac{m_\mathrm{tot}}{2E_\mathrm{orb}} \: ,
    \label{eq:par_from_sim}
\end{align}
where the ``a'' subscript refers to the apocenter. During the calculation of the mean anomaly $M$, for small $\delta M$ differences, due to numerical inaccuracies in determining the apocenter, we find that more accurate results can be obtained by directly calculating $M$ from the true anomaly $\varphi$, using
\begin{align}
    M =& \mathrm{atan2}\left( -\sqrt{1-e^2}\sin \varphi, -e-\cos \varphi \right) +\nonumber \\
    &+ \pi - e\frac{\sqrt{1-e^2}\sin \varphi}{1+e\cos \varphi} \: .
    \label{eq:MfromPhi}
\end{align}

Including the 1PN and 2PN terms introduces additional effects, such as
non-negligible apsidal precession, which modifies Eq.~\eqref{eq:MfromPhi}. We circumvent the complications by directly calculating the direction of the periastron using the Runge-Lenz vector, defined as
\begin{equation}
    \mathbf{A} = \mathbf{p}\times\mathbf{L} - \mu^2 m_\mathrm{tot}\frac{\mathbf{r}}{r} \: ,
\end{equation}
and adjusting for its direction when calculating the true anomaly and the mean anomaly afterwards. The components of the Runge-Lenz vector using polar coordinates are
\begin{align}
\mathbf{A} &= \mu^2 r^2\dot{\varphi}\begin{pmatrix}
  \dot{r}\sin\varphi + r\dot{\varphi}\cos\varphi\\
-\dot{r}\cos\varphi + r\dot{\varphi}\sin\varphi
\end{pmatrix}\nonumber \\
    &-\mu^2 m_\mathrm{tot} \begin{pmatrix}
    \cos\varphi\\
    \sin\varphi
\end{pmatrix} \: .
\end{align}
The 2-body motion with 1PN and 2PN interactions included can be projected into a quasi-Keplerian picture with modified orbital parameters, as well as orbital frequency \cite{Blanchet:2013haa}. In our analytic calculations we take into account the modification of $\omega_\mathrm{orb}$ due to PN interactions to the 2PN order as described in Ref.~\cite{Blanchet:2013haa} (see Eq.~(347a) in Section~10.2).

\begin{acknowledgements}
J. T. and P. K. acknowledge support by the National Research, Development and Innovation (NRDI) fund of Hungary, financed under the K 21 funding scheme, Project No. K 138277. This work was supported by the UK Science and Technology Facilities Council Grant Number ST/W000903/1 (to B.K.).
\end{acknowledgements}

\bibliography{EccNS}{}

\end{document}